\RequirePackage{fix-cm}
\documentclass[smallextended]{svjour3}       % onecolumn (second format)
\smartqed  % flush right qed marks, e.g. at end of proof
\usepackage{graphicx}

\usepackage[numbers]{natbib}
\usepackage{graphicx}%
\usepackage{multirow}%
\usepackage{amsmath,amssymb,amsfonts}%
\usepackage{mathrsfs}%
\usepackage[title]{appendix}%
\usepackage{xcolor}%
\usepackage{textcomp}%
\usepackage{manyfoot}%
\usepackage{booktabs}%
\usepackage{algorithm}%
\usepackage{algorithmicx}%
\usepackage{algpseudocode}%
\usepackage{listings}%

\usepackage{todonotes}
\usepackage{regexpatch}
\usepackage{doi}
\usepackage{xspace}

\usepackage{xcolor}        % Provides advanced color handling
\usepackage{colortbl}      % Adds color to tables
\usepackage{tcolorbox}
\usepackage{amsfonts}      % Provides access to additional math fonts
\usepackage{enumitem}      % Customizes lists and their layout
\usepackage{listings}
\usepackage{float}
\usepackage{xargs}
\RequirePackage[nolist,nohyperlinks]{acronym}
\begin{acronym}
    \acro{DSL}[DSL]{Domain-Specific Language}
    \acro{PCM}[PCM]{Palladio Component Model}
    \acro{XML}[XML]{Extensible Markup Language}
    \acro{EMF}[EMF]{Eclipse Modeling Framework~\cite{EMF_Book}}
    \acro{UML}[UML]{Unified Modeling Language}
    \acro{GDPR}[GDPR]{General Data Protection Regulation of the European Union}
    \acro{DAG}[DAG]{Directed Acyclic Graph}

    \acro{UML}[UML]{Unified Modeling Language~\cite{UML}}
    \acro{MDE}[MDE]{Model-Driven Engineering~\cite{Kent2002}}
    \acro{LSH}[LSH]{Locality-Sensitive Hashing~\cite{Pauleve2010}}
    \acro{LLM}[LLM]{Large Language Model}
    \acro{GPT}[GPT]{OpenAI's Generative Pre-trained Transformer}
    \acro{SysML}[SysML]{Systems Modeling Language}
    \acro{AST}[AST]{Abstract Syntax Tree}
    \acro{MOF}[MOF~\cite{MOF2016a}]{Meta Object Facility}
    \acro{OCL}[OCL]{Object Constraint Language}
    \acro{XMI}[XMI]{XML Metadata Interchange}
    \acro{URI}[URI]{Uniform Resource Identifier}
\end{acronym}

  {\list{}{\leftmargin=0.1in\rightmargin=0.1in}\item[]}%
  {\endlist}

\newcommand{\contribution}[1]{\hyperref[c#1]{\textbf{C#1}}}
\newcommand{\rqref}[1]{\hyperref[rq#1]{\textbf{RQ#1}}}
\newcommand{\qref}[1]{\hyperref[gqm-plan]{\textbf{Q#1}}}
\newcommand{\hyporef}[1]{\hyperref[hypo#1]{\textit{Hypothesis #1}}}
\newcommand{\obsref}[1]{\hyperref[obs#1]{\textit{Observation #1}}}
\newcommand{\probref}[1]{\hyperref[problem#1]{\textit{Problem #1}}}

\newcommand{\mossad}{{MOSSad}\xspace} % The original authors use small caps. Alternatives would be, e.g.,  caps lock

\newcommand{\summaryBox}[2]{\begin{tcolorbox}[colback=black!1!white,colframe=white!20!black,center,valign=top,halign=left,before skip=0.5cm,after skip=0.5cm,center title,sharp corners,boxrule=0.5pt,boxsep=1pt,width=\linewidth]\textbf{Answer to \qref{#1}:} \textit{#2}\end{tcolorbox}}

\newcommand{\B}[1]{\textbf{#1}} % more compact in tables

\begin{document}

\title{Evaluating Software Plagiarism Detection in the Age of AI
}
\subtitle{Automated Obfuscation and Lessons for Academic Integrity}

\author{Timur Sa\u{g}lam \and Larissa Schmid}

\institute{Timur Sa\u{g}lam \at
              KASTEL, Karlsruhe Institute of Technology (KIT), Germany \\
              \email{saglam@kit.edu}
           \and
           Larissa Schmid \at
              TCS, KTH Royal Institute of Technology, Sweden \\
              \email{lgschmid@kth.se}
}

\maketitle

\begin{abstract}
Plagiarism in programming assignments is a persistent issue in computer science education, increasingly complicated by the emergence of automated obfuscation attacks.
While software plagiarism detectors are widely used to identify suspicious similarities at scale and are resilient to simple obfuscation techniques, they are vulnerable to advanced obfuscation based on structural modification of program code that preserves the original program behavior. 
While different defense mechanisms have been proposed to increase resilience against these attacks, their current evaluation is limited to the scope of attacks used and lacks a comprehensive investigation regarding AI-based obfuscation.
In this paper, we investigate the resilience of these defense mechanisms against a broad range of automated obfuscation attacks, including both algorithmic and AI-generated methods, and for a wide variety of real-world datasets.
We evaluate the improvements of two defense mechanisms over the plagiarism detector JPlag across over four million pairwise program comparisons. 
Our results show significant improvements in detecting obfuscated plagiarism instances, and we observe an improved detection of AI-generated programs, even though the defense mechanisms are not designed for this use case.
Based on our findings, we provide an in-depth discussion of their broader implications for academic integrity and the role of AI in education.

\keywords{Source Code Plagiarism \and Plagiarism Detection \and Plagiarism Obfuscation \and Obfuscation Attack \and CS Education \and Generative AI}
\end{abstract}

\section{Introduction}

Plagiarism is a prevalent challenge in computer science education, facilitated by the ease of duplicating and modifying digital assignments \cite{Cosma2008, Murray2010, Le2013}.
Although students generally acknowledge plagiarism as academic misconduct, some will engage in it despite the threat of consequences~\cite{Sutton2014}.
Therefore, students are creative in \textit{obfuscating} their plagiarism to conceal the relation to its source~\cite{Pawelczak2018}. In the case of programming assignments, students commonly utilize techniques such as renaming, reordering, or restructuring~\cite{Novak2019, Karnalim2016}.
Plagiarism in programming assignments is particularly pronounced in beginner-level and mandatory courses, such as introductory programming courses~\cite{Park2003}.

While checking submissions for plagiarism manually is feasible for small course sizes, this quickly becomes infeasible for larger course sizes~\cite{Camp2017, Kustanto2009}  as the number of required pairwise comparisons grows quadratically -- reaching 1,225 comparisons for just 50 submissions. This results in the individual risk of detection decreasing with rising course sizes~\cite{Yan2018}.
In light of these issues, it is common for educators to use software plagiarism detection systems to uphold academic integrity for programming assignments~\cite{DevoreMcDonald2020}. These systems automate parts of the detection process and thus allow tackling the problem of plagiarism detection at scale. 
Thus, educators strongly rely on software plagiarism detectors to guide them in inspecting suspicious candidates. 
Plagiarism detectors analyze sets of programs to detect pairs with a suspiciously high degree of similarity~\cite{prechelt2000}.
However, assessing which suspicious candidates qualify as plagiarism is ultimately a human decision, given the underlying ethical considerations~\cite{Culwin2001, Weber2019}.
Overall, plagiarism detection systems help identify plagiarism instances and, when using such systems is communicated~\cite{Karnalim2022}, deter students from plagiarizing firsthand \cite{Braumoeller2001}.

%%% GENERAL CHALLENGE
Crucially, plagiarism detectors are only effective when defeating them takes more effort than completing the actual assignment~\cite{DevoreMcDonald2020}.
Yet, manually obfuscating a program successfully is tedious and requires understanding the underlying program, therefore requiring time and programming proficiency. 
Thus, a widespread assumption was that evading detection is not feasible for novice programmers as obfuscating the program requires more time than it takes to complete the actual assignments and requires a profound understanding of programming languages~\cite{Joy1999}.
However, this assumption has been broken with the recent rise of automated \textit{obfuscation attacks}~\cite{DevoreMcDonald2020, Foltynek2020, Biderman2022, Pawelczak2018} which require neither time nor programming proficiency to employ successfully. 
These \textit{obfuscation attacks} aim to avoid detection by strategically altering a plagiarized program, thus obscuring the relation to its original~\cite{Saglam2024b}: 
State-of-the-art detection approaches compare the structure of programs by identifying similarities between code fragments~\cite{Nichols2019}. Thus, most obfuscation attacks alter the structural properties of the program, ideally without affecting its behavior. 
Early automated attacks relied purely on algorithmic approaches, for example, via repeated statement insertion~\cite{DevoreMcDonald2020}. 
However, the challenge intensifies with the rise of generative artificial intelligence, especially \acp{LLM}~\cite{Daun2023}, making the obfuscation of plagiarism even more accessible with less effort than ever before~\cite{Khalil_Er_2023, Saglam2024a}. 
While state-of-the-art detectors exhibit some obfuscation resilience to changes like retyping and lexical changes, this does not apply to all types of obfuscation attacks~\cite{DevoreMcDonald2020, Luo2017}.
Thus, automated obfuscation attacks present a significant challenge for today's plagiarism detection systems, as they must now contend with increasingly sophisticated obfuscation techniques that can evade detection while maintaining the original program's functionality.

%%% THIS PAPER - RESEARCH GAP 
In recent work~\cite{Saglam2024b, Saglam2025}, we proposed defense mechanisms tailored towards different obfuscation attacks. However, it is unclear how well they work against a broader range of attacks.
Token Sequence Normalization~\cite{Saglam2024b} explicitly targets dead code insertion and statement reordering. 
Subsequence Match Merging~\cite{Saglam2025} employs a heuristic to counteract any obfuscation attack that aims at interrupting the match found between two program codes.
While we show that these approaches are effective against the obfuscation attacks they are targeting individually, educators typically do not know which obfuscation attacks students employed, thus making it hard to select a specific appropriate defense mechanism. 
Ideally, a combination of these defense mechanisms can be used to provide greater resilience. 
However, it is currently unclear whether the different defense mechanisms can be combined to achieve this while not producing false-positive results due to the overapproximation of similarities. 
Moreover, with the steady improvements of \acp{LLM}, AI-based obfuscation attacks become more and more feasible.
However, the defense mechanisms have not been comprehensively tested against a broad spectrum of automated attacks, including both algorithmic and AI-based obfuscation techniques. 
Finally, their applicability to detecting AI-generated programs is yet to be assessed.

\subsection{Research Contributions}

In this paper, we investigate the resilience of software plagiarism detectors to different automated obfuscation attacks. 
As a first contribution (C1), we present a comprehensive evaluation of various automated obfuscation attacks, including both algorithmic and AI-based methods, also exploring the feasibility of using AI to generate programs. 
In addition to examining defense mechanisms on their own, we also explore their combined use. 
As a second contribution (C2), we complement our technical findings with a detailed discussion of their broader implications -- not only for improving software plagiarism detection, but also for issues related to academic integrity and the role of AI in education.

\subsection{Evaluation and Results}
We conducted a comprehensive empirical evaluation to demonstrate the effectiveness of defense mechanisms against obfuscation attacks.
Over the entirety of this evaluation, we analyze over \textit{4 million data points}, each representing a pairwise comparison of two programs. Our datasets comprise over 14,000 files with over a million lines of code.

We evaluate the defense mechanisms with a wide range of real-world datasets~\cite{paiva2023, Ljubovic2020a, Saglam2024b} from different university courses. These courses range from mandatory undergraduate courses to master's-level elective courses. Furthermore, they contain different-sized programs, thus representing typical use cases for software plagiarism detection.
In our evaluation, we employ a total of \textit{five} different obfuscation techniques for the plagiarism instance.
We use both algorithmic and AI-based obfuscation and use existing obfuscation tools~\cite{gpt4, DevoreMcDonald2020}.

We demonstrate that the defense mechanisms offer broad obfuscation resilience across diverse datasets and attack types, thus significantly advancing resilience against automated obfuscation attacks for programming assignments.
Notably, we achieved a median similarity difference increase of up to 99.65 percentage points against semantic-preserving insertion-based obfuscation. We also show substantial improvements against refactoring-based attacks (up to 22 percentage points). While resilience against AI-based obfuscation was comparatively lower (up to 19 percentage points), we still observe improved detection rates, including a notable 8.92 percentage point increase in identifying AI-generated programs, even though the defense mechanisms are not designed for this use case.
These findings underscore the effectiveness of current defense mechanisms in defending against a wide range of obfuscation attacks, allowing for resilient source code plagiarism detection.

\subsection{Outline}
The remainder of this paper is structured as follows. First, \autoref{sec:two} introduces the foundations of automated obfuscation attacks and their impact on token-based plagiarism detection. In \autoref{sec:three}, we present the defense mechanisms designed to counter these attacks, which we evaluate in this paper. Next, \autoref{sec:four} outlines our evaluation methodology, followed by \autoref{sec:five}, which reports the results across various datasets and obfuscation strategies. We discuss threats to validity in \autoref{sec:six}, and provide a broader discussion of implications and insights in \autoref{sec:seven}. Finally, \autoref{sec:eight} reviews related work, and \autoref{sec:nine} concludes.

\section{Automated Obfuscation Attacks}\label{sec:two}
Students often attempt to conceal plagiarism by obfuscating its origin~\cite{Joy1999, Novak2020, Karnalim2016, Pawelczak2018}. Since cosmetic changes alone (e.g., lexical edits) are insufficient against structural comparison~\cite{Nichols2019}, they increasingly alter program structure while preserving its behavior. Common strategies include inserting statements, refactoring control structures~\cite{Karnalim2016}, or simplifying, combining, and splitting code fragments~\cite{Novak2019}.
These techniques, however, are neither new nor especially worrying, as manual obfuscation is tedious, error-prone, and requires understanding the original program to be plagiarized~\cite{Joy1999}.
\textit{Automated} obfuscation attacks, however, introduce a paradigm shift. Automated obfuscation is both faster and more effective than manual obfuscation.

\begin{figure}
    \centering
    \includegraphics[width=0.98\linewidth]{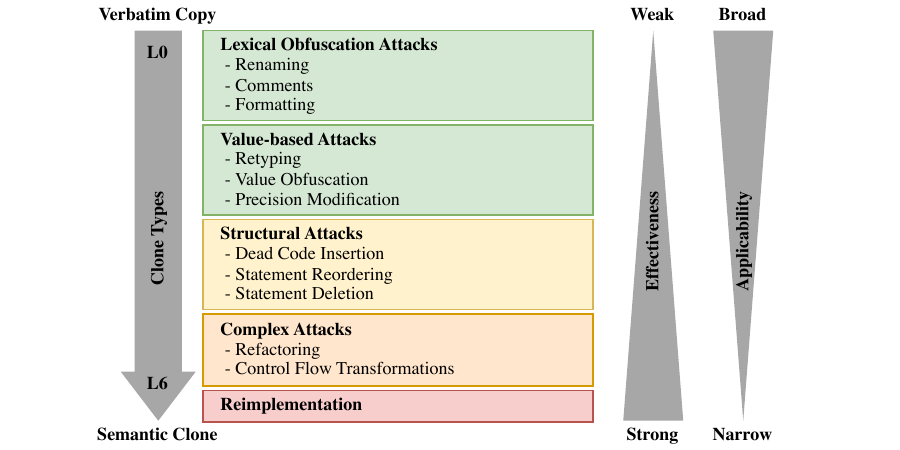}
    \caption[Categorization of Obfuscation Attack Types]{Categorization of obfuscation attacks corresponding to the clone types by \citet{Karnalim2016} based on \cite{Faidhi1987}, illustrating the effectiveness and applicability corresponding to the obfuscation complexity. Based on \cite{Saglam2025}.}
    \label{fig:clone-types}
\end{figure}

All automated obfuscation attacks targeting software plagiarism detectors -- whether manual, algorithmic, or AI-based -- are based on a single underlying principle: avoiding detection by strategically altering a plagiarized program, thus obscuring the relation to its original~\cite{Saglam2024b}. As state-of-the-art detection approaches compare the structure of programs by identifying similarities between code fragments~\cite{Nichols2019}, obfuscation attacks try to alter the structural properties of the program, ideally without affecting its behavior~\cite{Novak2020, Karnalim2016, Pawelczak2018}. Their intended outcome is to disrupt the matching of fragments between programs, thus leading to a reduced similarity score~\cite{DevoreMcDonald2020}. Specifically, the goal is to prevent the detector from matching fragments above the specified match length cut-off threshold. This can be achieved by breaking up matching code fragments into shorter sub-fragments.
However, to impact the detection quality of a software plagiarism detector, the obfuscation must affect the linearized program representation of the detector, which in the case of token-based approaches is the token sequence~\cite{Saglam2024b}. Consequently, modifications to the program code that do not affect the internal program representation are inherently ineffective. For example, renaming program elements does not affect token-based approaches, as names are omitted during the tokenization~\cite{prechelt2000, Saglam2024a} (see \autoref{fig:clone-types}).

\citet{DevoreMcDonald2020} present an automated attack based on repeated insertion of dead statements into an existing program.
This approach effectively deceives both JPlag~\cite{prechelt2002} and MOSS~\cite{MOSS}, reducing the calculated similarity between a plagiarism instance and its source below the average similarity of unrelated student solutions.
Similar attacks can be designed based on the automated application of refactoring operations~\cite{Maisch2024}. 
The rapid improvements in the field of generative artificial intelligence significantly exacerbate this problem~\cite{Lancaster2023}.
AI-powered tools can generate or alter source code~\cite{Daun2023} while requiring little manual effort and technical knowledge, making automated obfuscation more accessible than ever before \cite{Biderman2022, Khalil_Er_2023}. Tools like ChatGPT combine the capabilities of generative artificial intelligence with the approachable interface of a chatbot~\cite {Saglam2024a}, thus further reducing the entry barrier to using generative AI.
Essentially, automated obfuscation attacks make successfully evading plagiarism detection systems easier than ever.

\section{Defense Mechanisms}\label{sec:defense}\label{sec:three}
In the following, we present defense mechanisms against automated obfuscation attacks from our prior work~\cite{Saglam2024b, Saglam2025}.
These defense mechanisms are designed to provide broad resilience against automated obfuscation attacks by being largely language-independent, applicable across multiple programming languages, and agnostic to the underlying detection system, making them suitable for integration into any state-of-the-art, token-based detector such as MOSS, JPlag, or Dolos.
Token-based plagiarism detectors are inherently immune to lexical obfuscation, and usually also to data-based obfuscation (see \autoref{fig:clone-types}). Our defense mechanisms additionally provide resilience to structural and complex attacks.

\subsection{Token Sequence Normalization}
Token Sequence Normalization (TSN)~\cite{Saglam2024b} is a normalization-based defense mechanism designed to counter obfuscation attacks based on dead code insertion or statement reordering (\textit{structural attacks} in \autoref{fig:clone-types}). It uses a \textit{Token Normalization Graph} (TNG), a graph-based abstraction similar to program dependence graphs that captures semantic interdependencies between tokens. The normalization process begins by enriching the token sequence with language-independent semantic information. From this enriched sequence, a TNG is constructed to represent a partial ordering over tokens, abstracting away from the original code structure. The normalized token sequence is then generated from this graph by removing dead nodes and reverting reordered code via topological sorting. This normalization is performed before the pairwise comparison step in the plagiarism detection pipeline, allowing the detector to virtually de-obfuscate plagiarized code while preserving the scalability of token-based methods. The only language-specific component of this approach is the extraction of semantic information required for enrichment, which is consistent with the language-dependent nature of tokenization itself and does not introduce additional constraints on the detection system.

\subsection{Subsequence Match Merging}
Subsequence Match Merging (SMM)~\cite{Saglam2025} is a defense mechanism designed to counter obfuscation in an attack-independent and language-independent manner. Thus, it covers any of the categories presented in \autoref{fig:clone-types}. It is based on the observation that all effective obfuscation attacks must disrupt the matching of code fragments by breaking up the internal linearized program representation of detectors. SMM operates on these internal representations by heuristically merging neighboring fragment matches in pairs of programs.
This process is applied iteratively, subsuming gaps caused by obfuscation until no more neighboring matches can be merged.
SMM thus restores the continuity of matches, which reverts the effects of the obfuscation attack without significantly increasing the false positive rate.
Crucially, the approach is entirely language-independent and agnostic to the type of obfuscation, as it does not rely on the semantics of the internal representation.

\subsection{Combination of Both}
As TSN and SMM operate during different steps of the detection pipeline, they are complementary and can be combined. In our evaluation, we explore a hybrid defense strategy that applies TSN as a pre-processing step after parsing the input programs, followed by SMM as a post-processing step after computing matching subsequences. The rationale behind this combination is twofold.
  TSN provides strong resilience to insertion-based obfuscation, which is one of the easiest and most effective obfuscation attacks.
  SMM provides broad resilience against a range of obfuscation attacks, as its heuristic nature avoids making assumptions on the specifics of an obfuscation attack.
This layered approach is expected to offer broad resilience, combining the benefits of both approaches.

\section{Evaluation Methodology}\label{cha:methodology}\label{sec:four}
This section outlines the methodology used to evaluate the effectiveness of the proposed defense mechanisms regarding obfuscation resilience and detection quality.
We evaluate the aforementioned defense mechanisms regarding obfuscation resilience with the plagiarism detector JPlag as the baseline, as it is not only considered state-of-the-art~\cite{Aniceto2021} but also the most referenced and compared approach~\cite{Novak2019}. 
We use real-world datasets from different university courses.
We employ a total of four different obfuscation techniques for the plagiarism instances:
Dead code insertion, automated refactoring, AI-based obfuscation, and AI-based generation.
Over the entirety of this evaluation, we analyze over \textit{4.1 million data points}, each representing a similarity value of a pairwise comparison of two programs. The datasets sum up to over 14,000 files with around a million source lines of code.

In our evaluation, we analyze software plagiarism detectors for the purpose of evaluating their resilience with respect to automated obfuscation techniques in the context of computer science education. In this context, we investigate the following evaluation questions:

\begin{description}\label{main-gqm}\label{gqm-plan}
    \item[Q1] To what degree do defense mechanisms affect the similarity scores of unrelated programs?
    \item[Q2] What degree of resilience do defense mechanisms achieve against insertion-based obfuscation?
    \item[Q3] What degree of resilience do defense mechanisms achieve against refactoring-based obfuscation?
    \item[Q4] What degree of resilience do defense mechanisms achieve against AI-based obfuscation?
    \item[Q5] How well can we distinguish AI-generated from human programs?
    \item[Q6] What impact do defense mechanisms have on threshold-based plagiarism generators?
\end{description}
\noindent
For \textit{Q1}, we examine the similarity values for pairs of original programs as a metric. 
For \textit{Q2} to \textit{Q5}, we look at the similarity value differences between plagiarized and original programs and conduct comprehensive statistical tests by computing the statistical significance (p-values) as well as the practical significance (effect size) of these differences. 
To answer \textit{Q6}, we measure the difference in the runtime of the plagiarism generator and the difference in the number of inserted lines in the plagiarism instance. 

In the following, we outline the similarity metrics and statistical measures in detail. Next, we discuss our choice of baseline. We then describe the datasets we used. Finally, we explain obfuscation attacks used for obfuscation purposes.

\subsection{Similarity Metrics}\label{sec:metrics}

As plagiarism detection systems compute similarity scores between program pairs, these scores serve as the primary basis for identifying suspicious cases. In practice, similarity scores guide which candidates are reviewed first, as no objective indicator alone can confirm plagiarism. Detection tools typically provide similarity distributions and ranked lists of pairs -- both derived from similarity scores -- to support human inspection. The detailed visualization of matched code fragments is usually consulted only after identifying high-similarity candidates.
Evaluating detection quality requires distinguishing between different types of program pairs, each requiring separate analysis of the detector's similarity scores:
\begin{enumerate}[topsep=3pt]
\item \textit{Original Pair}: Two original programs developed independently of each other, without shared origin.
\item \textit{Plagiarism-To-Source Pair}: A plagiarism instance and its source program.
\end{enumerate}
To clearly distinguish plagiarism from unrelated programs during human inspection, plagiarism pairs must have high similarity scores~\cite{Saglam2024b}, while unrelated pairs must have low scores. Ideally, there is no overlap, with plagiarism pairs always showing higher similarity. However, in practice, overlap occurs as changes, especially obfuscation techniques, can reduce the similarity between a plagiarized program and its source. Thus, the \textit{difference} in similarity between plagiarism and unrelated pairs is crucial to measure detection quality.

A common anti-pattern in existing works is evaluating plagiarism detectors using a fixed similarity threshold where scores above count as successful detections, and those below do not. 
While this simplifies deriving precision, recall, and F1 scores, this approach is \textit{fundamentally flawed}, as the threshold can arbitrarily influence results and it can be tuned to favor one approach.
Since no universal threshold fits all datasets, thresholds are chosen arbitrarily. Due to varying similarity distributions for different datasets, they can only be set after seeing the results, thus introducing a strong bias.
A threshold-based evaluation reduces plagiarism detection to a binary classification problem, which is insufficient due to the mentioned problem of overlap.
Fundamentally, it measures only whether plagiarism is detected (and only according to some arbitrary criterion), not how well it is detected.

\begin{figure}[ht]
    \centering
    \begin{minipage}[t]{0.49\linewidth}
        \centering
        \includegraphics[width=\linewidth, trim=2.9cm 0cm 2.9cm 0cm, clip]{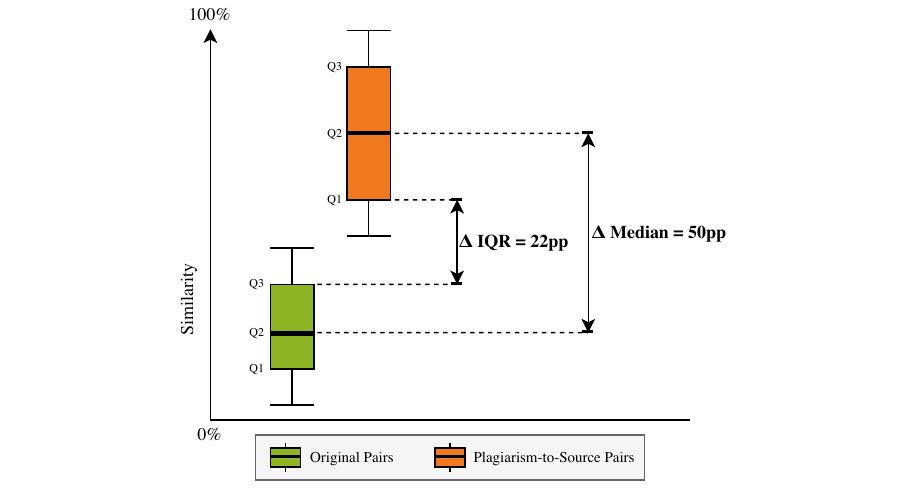}
        \caption{Difference metrics comparing the separation between plagiarism instances and original student solutions based on the median and interquartile range (IQR) distances measured in percentage points.}
        \label{fig:diff-metrics}
    \end{minipage}
    \hfill
    \begin{minipage}[t]{0.49\linewidth}
        \centering
        \includegraphics[width=\linewidth, trim=2.9cm 0cm 2.9cm 0cm, clip]{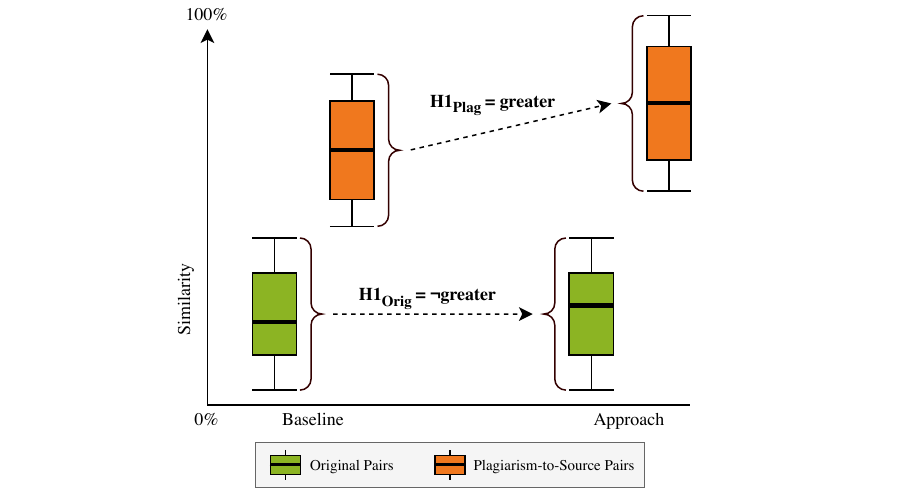}
        \caption{Visualization of alternative hypotheses ($H1$) for one-sided significance tests between plagiarism and original pairs. For plagiarism, the shift should be significantly greater ($H1_{Plag}$), while for originals, it should not ($H1_{Orig}$).}
        \label{fig:sig-metrics}
    \end{minipage}
\end{figure}

For these reasons, we focus on the difference in similarity between plagiarism and non-plagiarism pairs. The larger this difference, the easier it is to detect plagiarism effectively. 
Thus, we measure to \textit{what extent} a detection approach can produce such a difference between these pairs. This avoids overabstracting the problem into a binary classification.
Varying statistical measures can be used to calculate the differences between these types of pairs. 
We use measures of central tendency like the mean ($\mu$) or median ($Q_2$) and measures of spread like the difference between the interquartile ranges ($Q_1$ to $Q_3$).
\autoref{fig:diff-metrics} shows an \textbf{example} of the median difference ($\Delta median$) and interquartile range distance ($\Delta IQR$).
Each metric highlights different data properties: $\Delta median$ resists outliers more than $\Delta mean$, while $\Delta IQR$ indicates how well the main value ranges are separated. A $\Delta IQR$ above zero signifies that at least 75\% of values in both sets do not overlap.

We also employ statistical tests to assess statistical and practical significance. To that end we conduct one-sided Wilcoxon signed-rank tests to compare improvements in detection quality between approaches.
Different hypotheses apply depending on the pair type, as shown in \autoref{fig:sig-metrics}. For plagiarism pairs, we test if they show a significant location shift ($H1_{Plag}$), i.e., higher scores. For original pairs, we test that no significant shift occurs ($H1_{Orig}$).
For the practical significance, we use \textit{Cliff's delta} $\delta$~\cite{Cliff1993} as an effect size measure, as we deal with non-normal distributions and paired data.
Although a paired version of \textit{Cohen's d}~\cite{Cohen1988} exists, it is sensitive to outliers and only mildly robust to non-normality, making it unsuitable here. In contrast, while not ideal for paired data, \textit{Cliff's delta} $\delta$ remains useful for rank-based comparisons, offering robustness to non-normality and variance differences.
There are no established categories to interpret the resulting $\delta$ values. Thus, we base our interpretation on the derived categories by \citet{Romano2006} based on \textit{Cohens d}:

\small
\begin{equation}
    \delta \, \textit{Interpretation}=
    \begin{cases}
        \text{Negligible} & \text{if } 0\ \ \ \ \ \ \ \leq \lvert \delta \rvert < 0.147 \\
        \text{Small} & \text{if } 0.147  \leq \lvert \delta \rvert < 0.33 \\
        \text{Medium} & \text{if } 0.33 \ \ \leq \lvert \delta \rvert < 0.474 \\
        \text{Large} & \text{if } 0.474  \leq \lvert \delta \rvert < 0.7 \\
        \text{Very Large} & \text{if } 0.7 \ \ \ \ \leq \lvert \delta \rvert \leq 1
    \end{cases}
\end{equation}
\normalsize

Note that a negative effect size suggests an adverse interpretation, e.g., that the comparison group is greater than the target group. 

\subsection{Choice of Baselines}\label{sec:baselines}
For the evaluation of programming assignments, we utilize JPlag as our baseline. JPlag is not only regarded as a state-of-the-art tool~\cite{Aniceto2021, Novak2019} but also stands out as one of the most frequently referenced approaches and the most compared approach in the literature~\cite{Novak2019}. Its widespread use in practice and scientific literature makes it an ideal standard for assessing programming-based plagiarism.

While MOSS is widely used~\cite{Novak2019}, we excluded it as a baseline for four reasons: (1) it only returns a subset of similarity values, skewing comparisons; (2) it is closed-source and cannot be extended with our defense mechanisms; (3) it requires sending data to U.S.-based servers, conflicting with \ac{GDPR}; and (4) it imposes strict usage limits, making large-scale evaluation infeasible. We also excluded Dolos due to its limited adoption, and lack of support for multi-file programs. 
Multi-file programs are common in programming assignments, making it essential to evaluate plagiarism detection systems on such datasets.
Nonetheless, both MOSS and Dolos are token-based and equally vulnerable to obfuscation attacks~\cite{DevoreMcDonald2020, Saglam2024c}, underscoring the need for improved defenses.

\begin{figure}[b]
    \centering
    \includegraphics[width=0.99\linewidth]{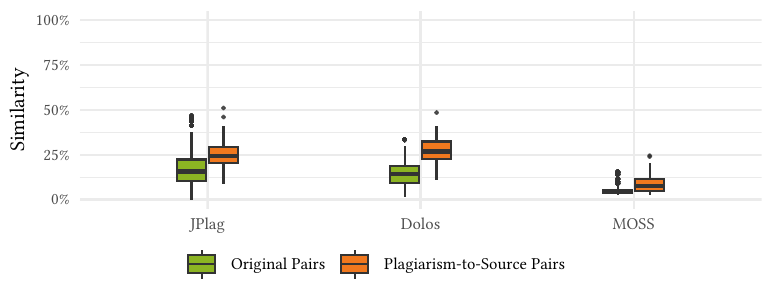}
    \caption[Obfuscation Vulnerability of Different Detection Systems]{Obfuscation vulnerability illustrated for the three token-based approaches JPlag, Dolos, and MOSS with the dataset PROGpedia-19~\cite{paiva2023} and plagiarism instances automatically obfuscated via statement insertion.}
    \label{fig:vuln}
\end{figure}
As an example, \autoref{fig:vuln} shows the results of JPlag, MOSS, and Dolos on a simple dataset using insertion-based obfuscation. All three tools yield low similarity scores for the plagiarism instances, causing overlap with unrelated programs. Note that MOSS omits many lower similarity values by design, and the dataset includes only small, single-file programs compatible with Dolos.

\subsection{Datasets}\label{sec:datasets}

We used a total of six real-world datasets; four are publicly available, and two are internal. Since public datasets are limited in both size and number, we supplement them with internal datasets.
All datasets come from an educational setting but stem from different courses and assignment types.
First, we used two tasks from the publicly available collection \textit{PROGpedia}~\cite{paiva2023}.
Here, Task 19 covers the design of a graph data structure and a depth-first search to analyze a social network.
Task 56 concerns minimum spanning trees using Prim's algorithm. Both datasets contain small Java programs.
For both datasets, we used only syntactically and semantically correct solutions and the latest version of each program.

Next, we used the \textit{TicTacToe} dataset~\cite{Saglam2024b}, which contains command-line-based Java implementations of the paper-and-pencil game TicTacToe. 
This dataset is from an introductory programming class at KIT, specifically from a weekly assignment.
This dataset contains many programs, each of which is medium-sized.
We also used the \textit{BoardGame} dataset~\cite{Saglam2024b}. This assignment is from the same course as the \textit{TicTacToe} dataset. However, it is the final project of the course. Here, the task is also a command-line-based game; however, this time, it is a comprehensive board game. Thus, it contains very large programs.

Finally, we used two tasks from the publicly available homework dataset by \citet{Ljubovic2020a}.
While both tasks contain C++ programs, one pertains to managing student and laptop records within a university setting, whereas the other requires implementing a Fourier series.
To prepare the datasets for our evaluation, we removed all solutions that did not compile, as JPlag requires valid input programs.
We also removed all human plagiarism (if present) based on the labeling provided by the datasets. If no labeling was present, we removed verbatim copies.
This notably reduces the size of some datasets.
Consequently, we obtained the six datasets listed in \autoref{tab:prog-datasets}.

\begin{table}[b]
    \centering
    \setlength{\tabcolsep}{5.8pt}
    \begin{tabular}{lrrrrr}
        \toprule
        {Dataset Name} & {\# Programs} & {Mean LOC} & Total LOC & Language & Source \\
        \midrule
        PROGpedia Task 19 & 27 & 131 & 3.5K & Java & \cite{paiva2023}\\
        PROGpedia Task 56 & 28 & 85 & 2.4K & Java & \cite{paiva2023}\\
        TicTacToe & 626 & 236 & 148K & Java & \scriptsize{internal}\\
        BoardGame & 434 & 1529 & 663K & Java & \scriptsize{internal}\\
        Homework Task 1 & 59 & 282 & 17K & C/C++ & \cite{Ljubovic2020a}\\
        Homework Task 5 & 18 & 123 & 2.2K & C/C++ & \cite{Ljubovic2020a}\\
        \bottomrule
    \end{tabular}
    \caption[Evaluation Datasets]{Programming assignment datasets used for the evaluation with the number of included programs, mean size in lines of code (LOC) excluding comments, the programming language, and source of the dataset.}
    \label{tab:prog-datasets}
\end{table}

\subsection{Obfuscation Attacks} \label{sec:methodology:attacks}
We evaluate four automated obfuscation attacks, two algorithmic and two AI-based. For ethical reasons, we briefly discuss these attacks without revealing details to avoid encouraging their use.
 
The first algorithmic obfuscation attack is the insertion of dead statements.
For this, we employ two different tools. The first one is \mossad~\cite{DevoreMcDonald2020}. As previously discussed, it is indeterministic and operates threshold-based. The second is \textit{PlagGen}~\cite{Broedel2023}, which is similar to \mossad but is deterministic and exhaustive. In both cases, the statement insertion uses statements from the original program and a pool of pre-defined statements. Furthermore, both ensure that the inserted statements do not change the behavior of the programs. Thus, this obfuscation attack is semantic-preserving. We use PlagGen for Java and \mossad for C++, as these are the languages supported by each tool.
Second, we employ the refactoring-based obfuscation attack by \citet{Maisch2024}, which leverages \textit{Spoon}~\cite{Pawlak2006} and automatically applies semantic-preserving refactoring operations at random positions at the AST level to obfuscate a program.
In detail, the refactoring operations include optional wrapping, extracting expressions as new variables, introducing constant container classes and extraction of constants, swapping if-else-statements and inverting the corresponding conditions, inserting methods and constructors, and introducing access methods for existing fields.
As the behavior of the programs is not changed, this obfuscation attack is also semantic preserving.
This implementation of the obfuscation attack only supports Java programs, so we only use four of the six datasets with this obfuscation attack.

For the AI-based obfuscation attacks, we exploit OpenAI's GPT-4 for automated plagiarism, which is currently the state-of-the-art LLM.
There are generally two ways of using generative AI to \textit{cheat} for programming assignments:
\textit{AI-based obfuscation}, where the adversary provides an AI model with a pre-existing program and tasks it to generate an obfuscated version.
\textit{AI-based generation}, where the adversary uses the assignment's description to generate a program from scratch via an AI model.
We employ AI-based obfuscation as a third obfuscation attack alongside both algorithmic ones. We use fifteen different prompts, mimicking how students would ask GPT to obfuscate their plagiarism.
The prompts range from requesting minor structural changes to requesting a reimplemented version of the original program. As for this attack, the programs need to be sent to the OpenAI GPT server; we did not use it for the BoardGame dataset due to its sensitive nature.
Finally, we use full generation as the final obfuscation. However, we can only employ it for the TicTacToe dataset, as we require the full assignment description and test cases to test for the expected behavior.
AI-based obfuscation is a semantic agnostic attack. While the prompts contain instructions to preserve the program behavior, there are generally no guarantees that the changes proposed by GPT-4 conform to these instructions. 
Similarly, for AI-based generation, there is no guarantee that the programs fully implement all details requested by the task.
%%%%%%%
In sum, we use the following four techniques to create 787 plagiarized programs (see \autoref{tab:plagiate} for details):
\begin{enumerate}[noitemsep]
    
 \item \textbf{Insertion-based Obfuscation} (semantic-preserving): Inserting dead statements into the program (PlagGen~\cite{Broedel2023} for Java and \mossad~\cite{DevoreMcDonald2020} for C/C++).
 \item \textbf{Refactoring-based Obfuscation} (semantic-preserving): Applying a variety of semantic-preserving refactoring operations, for example, transformations of control structures, field access, and method granularity~\cite{Maisch2024}.
 \item \textbf{AI-based Obfuscation} (sematic-agnostic): We obfuscate human solutions with GPT-4~\cite{gpt4} based on 15 varying prompts requesting structural changes.
 \item \textbf{AI-based Generation} (sematic-agnostic): We fully generate AI-based solutions with GPT-4~\cite{gpt4} based on only the textual task description of the assignment.
\end{enumerate}
\begin{table}[b]
	\centering
    \setlength{\tabcolsep}{1.4pt}
	\begin{tabular}{lcccccc}
		\toprule
		Obfuscation Attack Type  & PROGp.-19 & PROGp.-56 & Homew.-1 & Homew.-5 & TTT & BoardGame\\
		%Type                     & PROGpedia-19 & PROGpedia-56 & Homework-1 & Homework-5 & TicTacToe \\
		\midrule
		Insertion-based Obf. & 27    & 28    & 59   & 17   & 50 & 20 \\
		Alteration-based Obf.   & 27    & 28    & 59   & 17   & 50 & 20 \\
        Refactoring-based Obf.   & 27    & 28    & -   & -  & 50 & 20 \\
		AI-based Obf. & 75    & 75    & 75   & 75   & 75 & - \\
		AI-based Generation      & 50     & 50     & -    & -    & 50 & -\\
		\bottomrule
	\end{tabular}
    \caption[Obfuscation Attacks]{Overview on the number of plagiarized programs per dataset and obfuscation attack type (851 in total). Each of the 15 prompts is applied to 5 originals for the AI-based obfuscation.}
    \label{tab:plagiate}
\end{table}

\section{Evaluation Results}\label{cha:code-eval}\label{sec:five}
In the following, we provide the results of our evaluation, which demonstrate that the defense mechanisms offer broad obfuscation resilience across diverse datasets and attack types. We compare them to JPlag without any defense mechanisms as the baseline.
We provide a replication package for this evaluation~\cite{replication-package}.
The key findings from our comprehensive evaluation, which offer new insights beyond prior evaluations, can be summarized as follows:
    \textit{Insertion-based Obfuscation:} Combining both defense mechanisms provides improved resilience.
    \textit{Refactoring-based Obfuscation:} Token sequence normalization minimally impacts refactoring-based attacks, as expected. However, subsequence match merging significantly improves detection, and combining both mechanisms achieves enhanced separation of plagiarized and original programs. 
    \textit{GPT-4-based Obfuscation:} Token sequence normalization has no positive or negative impact on AI-based obfuscation. 
    Combining token sequence normalization and subsequence match merging shows significant improvements without drawbacks.
    \textit{GPT-4-generated Programs:} Despite the defense mechanisms not being tailored for AI-generated program detection, we observe significantly improved detection across datasets for subsequence match merging. 
    \textit{Threshold-based Plagiarism:} The defense mechanisms substantially increase the computational cost of threshold-based obfuscation, making such an obfuscation method more tedious and also easily detectable via metrics such as program size or number of tokens.

In summary, our evaluation demonstrates that the proposed defense mechanisms are highly effective across a range of automated obfuscation attacks.
The proposed defense mechanisms provide significantly (statistical and practical significance) improved obfuscation resilience without any practically significant change in false-positive rates.
In the following, we present detailed results for each attack type as outlined in \autoref{sec:methodology:attacks} individually.
The original, human-made programs of each dataset remain the same for all evaluation stages. Thus, we first discuss the effect of the defense mechanism on these unrelated programs.

% ------------------------------------------------------- $
\begin{table}
\centering
%\small
\setlength{\tabcolsep}{4pt}
\begin{tabular}{lrrrrrrrr}
  \toprule
ds & Variant & Pairs & $p$ & $W$ & $\delta$ & $\delta\,Int.$ & $\delta$ 95\% CI & n \\ 
  \midrule
    \multirow{3}{*}{\rotatebox[origin=c]{90}{Pp.-19}} & TSN & OP & 8.3e-05 & 13,137 & 0.005 & Negligible & [-0.08, 0.09] & 351 \\ 
   & SMM & OP & $<$ 1e-10 & 6,441 & 0.178 & Small & [0.09, 0.26] & 351 \\ 
   & Both & OP & $<$ 1e-10 & 26,133 & 0.185 & Small & [0.10, 0.27] & 351 \\ 
   \hline
  \multirow{3}{*}{\rotatebox[origin=c]{90}{Pp.-56}} & TSN & OP & 0.25 & 3,580 & -0.034 & Negligible & [-0.11, 0.04] & 378 \\ 
   & SMM & OP & $<$ 1e-10 & 4,753 & 0.154 & Small & [0.08, 0.23] & 378 \\ 
   & Both & OP & $<$ 1e-10 & 14,018 & 0.148 & Negligible & [0.07, 0.22] & 378 \\ 
  \hline
   \multirow{3}{*}{\rotatebox[origin=c]{90}{TTT}} & TSN & OP & $<$ 1e-10 & 1,256,914,634 & 0.009 & Negligible & [0.01, 0.01] & 188,805 \\ 
   & SMM & OP & $<$ 1e-10 & 3,346,969,836 & 0.177 & Small & [0.17, 0.18] & 188,805 \\ 
   & Both & OP & $<$ 1e-10 & 6,099,071,135 & 0.186 & Small & [0.18, 0.19] & 188,805 \\ 
   \hline
  \multirow{3}{*}{\rotatebox[origin=c]{90}{BG}} & TSN & OP & $<$ 1e-10 & 1,379,905,112 & 0.021 & Negligible & [0.01, 0.03] & 67,161 \\ 
   & SMM & OP & $<$ 1e-10 & 1,637,665,065 & 0.104 & Negligible & [0.10, 0.11] & 67,161 \\ 
   & Both & OP & $<$ 1e-10 & 2,069,412,722 & 0.124 & Negligible & [0.12, 0.13] & 67,161 \\ 
   \hline
  \multirow{3}{*}{\rotatebox[origin=c]{90}{Hw.-1}} & TSN & OP & 1 & 378,825 & -0.069 & Negligible & [-0.11, -0.03] & 1,711 \\ 
   & SMM & OP & $<$ 1e-10 & 557,040 & 0.161 & Small & [0.12, 0.20] & 1,711 \\ 
   & Both & OP & $<$ 1e-10 & 656,971 & 0.106 & Negligible & [0.07, 0.14] & 1,711 \\ 
   \hline
  \multirow{3}{*}{\rotatebox[origin=c]{90}{Hw.-5}} & TSN & OP & 0.58 & 2,229 & -0.058 & Negligible & [-0.19, 0.07] & 153 \\ 
   & SMM & OP & 3.9e-10 & 1,275 & 0.111 & Negligible & [-0.02, 0.24] & 153 \\ 
   & Both & OP & 0.015 & 3,153 & 0.062 & Negligible & [-0.07, 0.19] & 153 \\ 
   \bottomrule
\end{tabular}
\caption[Statistical Tests: Unrelated Pairs]{One-sided Wilcoxon signed-rank test results for \textbf{unrelated, student-made programs} regarding the potential adverse effects of our defense mechanisms compared to the baseline (sig. level of $\alpha=0.01$, alternative hypothesis $H1=greater$, test statistic $W$, effect size via Cliff's delta $\delta$, its interpretation $\delta\,Int.$, its confidence interval $CI$, and the sample size $n$). Note that high $p$ and low $\delta$ are desirable, as original pairs should \textit{not} be $greater$.} 
\label{tab:to-base-original}
\end{table}
 
% ------------------------------------------------------- $

\subsection{Effect on Unrelated Programs}\label{sec:eval-unrel}
Effective plagiarism detection requires not only high similarity for plagiarism pairs but also minimal impact on unrelated, original programs. Our results show that the effect on such unrelated programs is negligible, indicating no meaningful increase in false positives.

\autoref{tab:to-base-original} presents statistical test results for the original pairs, comparing defense mechanisms to the baseline.
Token sequence normalization has virtually no impact on unrelated programs: median similarity changes range from $-1.23$ to $+0.28$ percentage points across datasets, with only one statistically significant change (PROGpedia-19), which remains insignificant due to the negligible effect size.
Subsequence match merging yields small median increases ($+0.78$ to $+6.59$), statistically significant in four datasets, yet with negligible to small effect sizes and thus little to no practical significance.

When both mechanisms are combined, results are comparable to subsequence match merging alone, with median increases from $+0.91$ to $+6.75$. While statistically significant in three datasets, effect sizes remain negligible to small, confirming little to no practical significance for the impact on unrelated programs.

\summaryBox{1}{The defense mechanisms have a negligible effect on unrelated programs, meaning their impact on the false positive rate is both practically and, in some cases, statistically insignificant.}

\subsection{Insertion-based Obfuscation}\label{sec:eval-insert}
\begin{figure}
\centering
\includegraphics[width=\linewidth, trim=0 5pt 0 0, clip]{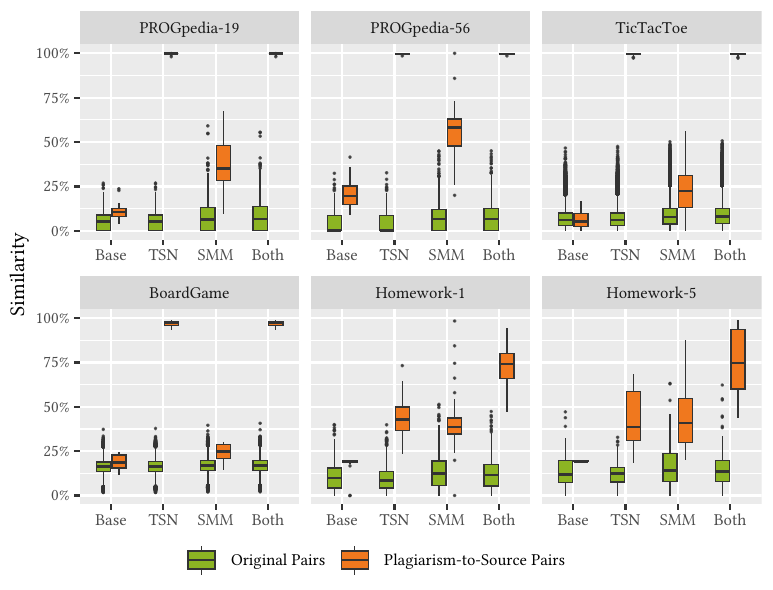}
\caption[Evaluation Results: Insertion-based Obfuscation]{Similarity scores for original program pairs and \textbf{insertion-based plagiarism} pairs. Ideally, plagiarism pairs exhibit high similarity, while original pairs should exhibit low similarity.}
\label{fig:stage1-results}
\end{figure}

\autoref{fig:stage1-results} presents results for insertion-based obfuscation attacks, with corresponding statistical measures in \autoref{tab:diff-insert}.
As described earlier, we use \mossad~\cite{DevoreMcDonald2020} for C++ datasets (Homework-1 and Homework-5) and PlagGen~\cite{Broedel2023} for the others. The key difference lies in their termination strategies: \mossad uses a threshold-based approach, while PlagGen exhaustively inserts code in all possible positions.

\begin{table}
	\centering
    \setlength{\tabcolsep}{3pt}
	\begin{tabular}{lrrrrrrrr}
		\toprule
		Dataset                       & Variant & Median     & Mean      & $Q_1$      & $Q_3$      & $\Delta$ Mean & $\Delta$ Median & $\Delta$ IQR \\
		\midrule
		\multirow{4}{*}{PROGpedia-19} & Base     & 10.61      & 11.10     & 8.17       & 12.67      & 5.37          & 5.55            & -0.74        \\ 
		                              & TSN      & \B{100.00} & \B{99.87} & \B{100.00} & \B{100.00} & \B{94.12}     & \B{94.87}       & \B{91.06}    \\ 
		                              & SMM      & 35.10      & 37.12     & 28.36      & 47.94      & 27.97         & 28.74           & 15.16        \\ 
		                              & Both     & \B{100.00} & \B{99.87} & \B{100.00} & \B{100.00} & 90.66         & 93.43           & 86.12        \\ 
		\hline
		\multirow{4}{*}{PROGpedia-56} & Base     & 19.59      & 21.36     & 14.86      & 25.33      & 16.47         & 19.59           & 6.09         \\ 
		                              & TSN      & \B{99.65}  & \B{99.66} & \B{99.47}  & \B{100.00} & \B{95.14}     & \B{99.65}       & \B{90.82}    \\ 
		                              & SMM      & 58.08      & 56.70     & 47.75      & 62.91      & 48.44         & 51.49           & 35.56        \\ 
		                              & Both     & \B{99.65}  & \B{99.66} & \B{99.47}  & \B{100.00} & 91.54         & 92.91           & 86.85        \\ 
		\hline
		\multirow{4}{*}{TicTacToe}    & Base     & 5.29       & 6.36      & 2.57       & 9.66       & -0.49         & -0.78           & -7.39        \\ 
		                              & TSN      & \B{99.52}  & \B{99.42} & \B{99.18}  & \B{100.00} & \B{92.48}     & \B{93.37}       & \B{89.11}    \\ 
		                              & SMM      & 22.36      & 23.39     & 13.01      & 31.14      & 14.57         & 14.52           & 0.49         \\ 
		                              & Both     & \B{99.52}  & \B{99.42} & \B{99.18}  & \B{100.00} & 90.49         & 91.58           & 86.54        \\
        \hline
		\multirow{4}{*}{BoardGame}    & Base     & 18.72      & 18.66     & 15.49      & 22.85      & 2.48          & 2.53            & -3.49        \\ 
		                              & TSN      & \B{97.23}  & \B{96.82} & \B{95.82}  & \B{98.07}  & \B{80.49}     & \B{80.90}       & \B{76.69}    \\ 
		                              & SMM      & 24.97      & 24.12     & 20.81      & 28.65      & 7.18          & 8.01            & 1.06         \\ 
		                              & Both     & \B{97.23}  & \B{96.82} & \B{95.82}  & \B{98.07}  & 79.72         & 80.13           & 75.92        \\
		\hline
		\multirow{4}{*}{Homework-1}   & Base     & 19.45      & 18.36     & 18.86      & 19.90      & 7.97          & 9.57            & 3.27         \\ 
    		                           & TSN      & 42.83      & 43.58     & 36.57      & 49.84      & 34.21         & 34.19           & 22.92        \\ 
		                               & SMM      & 38.57      & 40.63     & 34.75      & 43.63      & 27.58         & 25.99           & 15.25        \\ 
		                               & Both     & \textbf{74.10}      & \textbf{73.32}     & \textbf{65.88}      & \textbf{80.09}      & \textbf{61.32}         & \textbf{62.48}           & \textbf{48.45}        \\ 
		\hline
		\multirow{4}{*}{Homework-5}   & Base     & 19.24      & 19.38     & 19.03      & 19.91      & 5.99          & 7.25            & -0.65        \\ 
		                               & TSN      & 38.63      & 42.05     & 30.90      & 58.65      & 30.39         & 26.36           & 15.06        \\ 
		                               & SMM      & 40.89      & 43.36     & 30.05      & 54.77      & 27.44         & 26.91           & 6.40         \\ 
		                               & Both     & \textbf{74.57}      & \textbf{75.21}     & \textbf{59.98}      & \textbf{93.46}      & \textbf{60.62}         & \textbf{61.22}           & \textbf{40.40}        \\  
		\bottomrule  
	\end{tabular}
	\caption[Evaluation Results: Insertion-based Obfuscation]{Statistical measures for plagiarism pairs and their differences ($\Delta$) from original pairs for \textbf{insertion-based obfuscation} (corresponds to \autoref{fig:stage1-results}). Higher values indicate better performance. Note that measures are expressed as percentages and their differences as percentage points. Highest values by a margin of 0.25 are marked in bold.}
	\label{tab:diff-insert}
\end{table}

% ------------------------------------------------------- $
\begin{table}%[h]
	\centering
	%\small
    \setlength{\tabcolsep}{4pt}
	\begin{tabular}{lrrrrrrrr}
		\toprule
		ds                              & Variant & Pairs & $p$     & $W$   & $\delta$ & $\delta\,Int.$ & $\delta$ 95\% CI & n  \\ 
		\midrule
		\multirow{3}{*}{{PROGpedia-19}} & TSN     & P2S   & 3e-06   & 378   & 1.000    & Very Large     & [1.00, 1.00]     & 27 \\ 
		                                & SMM     & P2S   & 3e-06   & 378   & 0.904    & Very Large     & [0.70, 0.97]     & 27 \\ 
		                                & Both    & P2S   & 3e-06   & 378   & 1.000    & Very Large     & [1.00, 1.00]     & 27 \\ 
		\hline
		\multirow{3}{*}{{PROGpedia-56}} & TSN     & P2S   & 2e-06   & 406   & 1.000    & Very Large     & [1.00, 1.00]     & 28 \\ 
		                                & SMM     & P2S   & 2e-06   & 406   & 0.939    & Very Large     & [0.79, 0.98]     & 28 \\ 
		                                & Both    & P2S   & 2e-06   & 406   & 1.000    & Very Large     & [1.00, 1.00]     & 28 \\ 
        \hline
		\multirow{3}{*}{{TicTacToe}}    & TSN     & P2S   & 3.9e-10 & 1,275 & 1.000    & Very Large     & [1.00, 1.00]     & 50 \\ 
		                                & SMM     & P2S   & 8.4e-10 & 1,176 & 0.766    & Very Large     & [0.60, 0.87]     & 50 \\ 
		                                & Both    & P2S   & 3.9e-10 & 1,275 & 1.000    & Very Large     & [1.00, 1.00]     & 50 \\ 
        \hline
		\multirow{3}{*}{{BoardGame}}    & TSN     & P2S   & 4.8e-05 & 210   & 1.000    & Very Large     & [1.00, 1.00]     & 20 \\ 
		                                & SMM     & P2S   & 4.8e-05 & 210   & 0.545    & Large          & [0.20, 0.77]     & 20 \\ 
		                                & Both    & P2S   & 4.8e-05 & 210   & 1.000    & Very Large     & [1.00, 1.00]     & 20 \\ 
		\hline
		\multirow{3}{*}{{Homework-1}}   & TSN     & P2S   & $<$ 1e-10 & 1,770 & 1.000    & Very Large     & [1.00, 1.00]     & 59 \\ 
		                                & SMM     & P2S   & $<$ 1e-10 & 1,653 & 0.955    & Very Large     & [0.82, 0.99]     & 59 \\ 
		                                & Both    & P2S   & $<$ 1e-10 & 1,770 & 1.000    & Very Large     & [1.00, 1.00]     & 59 \\  
		\hline
		\multirow{3}{*}{{Homework-5}}   & TSN     & P2S   & 1.3e-4  & 170   & 0.889    & Very Large     & [0.45, 0.98]     & 18 \\ 
		                                & SMM     & P2S   & 1.6e-4  & 153   & 0.969    & Very Large     & [0.83, 0.99]     & 18 \\ 
		                                & Both    & P2S   & 1.1e-4  & 171   & 1.000    & Very Large     & [0.99, 1.00]     & 18 \\ 
		\bottomrule
	\end{tabular}
	\caption[Statistical Tests: Insertion-based Obfuscation]{One-sided Wilcoxon signed-rank test results for \textbf{insertion-based obfuscation} regarding the improvement by our defense mechanism compared to baseline (sig. level of $\alpha=0.01$, alternative hypothesis $H1=greater$, test statistic $W$, effect size via Cliff's delta $\delta$, its interpretation $\delta\,Int.$, its 95 percent confidence interval $CI$, and the sample size $n$). For plagiarism-to-source pairs (P2S), low $p$ and high $\delta$ are desirable.} 
	\label{tab:to-base-insert}
\end{table}
 
% ------------------------------------------------------- $

\subsubsection*{Baseline}
\autoref{fig:stage1-results} shows the severe impact of insertion-based obfuscation on the baseline. Median similarity values for plagiarism pairs drop to between 5.29\% (TicTacToe) and 19.59\% (PROGpedia-56), leading to near-complete overlap with original pairs in all datasets except PROGpedia-56, which still shows substantial overlap.
In the TicTacToe dataset, median similarity for plagiarism pairs is even \textit{lower} than for original pairs. As detailed in \autoref{tab:diff-insert}, the median similarity difference between plagiarism and original pairs ranges from $-0.78$ percentage points (TicTacToe) to $+19.56$ (PROGpedia-56), with most datasets showing differences below ten points -- indicating little to no separation.
These results confirm that insertion-based obfuscation is highly effective against JPlag and similar token-based detectors, significantly hindering plagiarism detection in the absence of additional defenses.

\subsubsection*{Token Sequence Normalization}
Token sequence normalization is designed to counter insertion-based obfuscation, and the results in \autoref{fig:stage1-results} confirm its strong effectiveness. For the Java datasets, it renders JPlag effectively immune to such attacks.
Similarity values for plagiarism pairs increase significantly across all datasets. In the Java datasets, median similarities reach between 97.23\% (BoardGame) and 100.00\% (PROGpedia-19), eliminating overlap with original pairs and achieving full separation. In the C++ datasets, values rise to 38.63\% (Homework-5) and 42.83\% (Homework-1), with the remaining overlap confined to edge quartiles.
This improvement is also evident in the similarity differences between plagiarism and original pairs (\autoref{tab:diff-insert}): 80.09 to 99.65 percentage points for Java, and 26.36 to 34.21 for C++ datasets.
Statistical tests confirm both statistical and practical significance across all datasets (\autoref{tab:to-base-insert}), with low p-values and very large effect sizes.
Overall, token sequence normalization substantially improves obfuscation resilience and demonstrates the value of targeted defenses against insertion-based obfuscation.

\subsubsection*{Subsequence Match Merging}
As an attack-independent defense, subsequence match merging is not specifically tailored to insertion-based obfuscation, but still yields notable improvements over the baseline (\autoref{fig:stage1-results}).
Median similarity values for plagiarism pairs increase across all datasets, ranging from 22.36\% (TicTacToe) to 58.08\% (PROGpedia-56), with overlap mostly confined to quartile extremes. Corresponding median similarity \textit{differences} between plagiarism and original pairs (\autoref{tab:diff-insert}) range from 8.01 (BoardGame) to 51.49 percentage points (PROGpedia-56), indicating strong separation.
Statistical tests (\autoref{tab:to-base-insert}) confirm that improvements are both statistically and practically significant. P-values are low across all datasets, and effect sizes are very large, except for BoardGame, which shows a large effect.
While less effective than token sequence normalization, subsequence match merging still offers strong resilience against insertion-based obfuscation -- especially notable given its general-purpose nature.

\subsubsection*{Combination of Both}
Combining both defense mechanisms yields strong improvements over the baseline, rendering JPlag effectively immune to insertion-based attacks. For Java datasets, results are comparable to subsequence match merging alone, while C++ datasets show clear gains over either individual method.
We no longer observe any significant overlap between the plagiarism and original pairs, resulting in a clear separation between both types of pairs.
Median similarity \textit{differences} (\autoref{tab:diff-insert}) range from 61.22 (Homework-5) to 93.42 percentage points (PROGpedia-19), indicating substantial improvements across all datasets.
Statistical tests (\autoref{tab:to-base-insert}) confirm that these gains are both statistically and practically significant, with low p-values and very large effect sizes for all datasets.
In summary, the combination of both mechanisms provides robust resilience against insertion-based obfuscation, especially strengthening detection for C++ programs.

\summaryBox{2}{The defense mechanisms significantly increase the resilience against semantic preserving insertion-based obfuscation attacks. The median similarity differences increase, depending on the dataset, up to 99.65 percentage points, thus producing a complete separation of plagiarized and original programs. Thus, the degree of resilience effectively reflects near-immunity to insertion-based attacks.}

\subsection{Refactoring-based Obfuscation}\label{sec:eval-refactor}
\begin{figure}
\centering
\includegraphics[width=\linewidth, trim=0 5pt 0 0, clip]{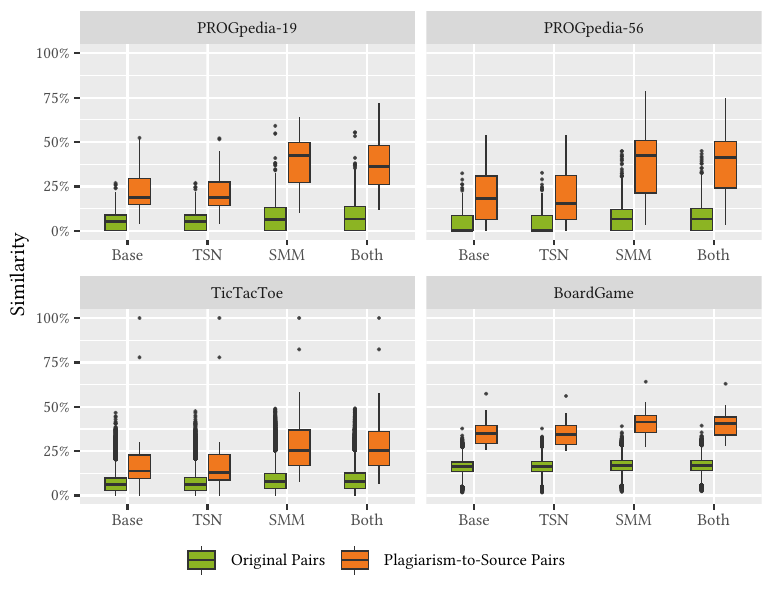}
\caption[Evaluation Results: Refactoring-based Obfuscation]{Similarity scores for original program pairs and \textbf{refactoring-based plagiarism} pairs. Ideally, plagiarism pairs exhibit high similarity, while original pairs should exhibit low similarity.}
\label{fig:stage2.5-results}
\end{figure}

\autoref{fig:stage2.5-results} presents results for refactoring-based obfuscation, which applies a mix of semantic-preserving refactorings at random positions in the parse tree of programs. Corresponding statistical measures are shown in \autoref{tab:diff-refactor}.
As the obfuscation tool~\cite{Maisch2024} supports only Java, this evaluation stage includes four of the six datasets.

\begin{table}
	\centering
    \setlength{\tabcolsep}{3pt}
	\begin{tabular}{lrrrrrrrr}
		\toprule
		Dataset      & Variant & Median    & Mean      & $Q_1$     & $Q_3$      & $\Delta$ Mean & $\Delta$ Median & $\Delta$ IQR \\ 			
		\midrule
		\multirow{4}{*}{PROGpedia-19} & Base & 18.82     & 23.21     & 14.74     & 29.64      & 17.48         & 13.76           & 5.83         \\ 
		 & TSN  & 18.82     & 22.48     & 14.43     & 27.53      & 16.73         & 13.70           & 5.49         \\ 
		 & SMM  & \B{42.29} & \B{38.63} & \B{27.35} & \B{49.59}  & \B{29.48}     & \B{35.93}       & \B{14.15}    \\ 
		 & Both & 36.26     & 37.15     & 26.23     & 48.21      & 27.93         & 29.69           & 12.35        \\ 
		\hline
		\multirow{4}{*}{PROGpedia-56} & Base & 18.20     & 21.55     & 6.42      & 30.97      & 16.66         & 18.20           & -2.34        \\ 
		 & TSN  & 15.47     & 20.85     & 6.43      & 31.10      & 16.34         & 15.47           & -2.22        \\ 
		 & SMM  & \B{42.62} & \B{37.02}     & 21.33     & \B{50.94}  & 28.76         & \B{36.02}       & 9.13         \\ 
		 & Both & 41.32     & \B{37.21} & \B{24.05} & 50.40      & \B{29.09}     & 34.58           & \B{11.44}    \\ 
		\hline
		\multirow{4}{*}{TicTacToe}    & Base & 13.90     & 17.86     & 9.49      & 22.84      & 11.01         & 7.83            & -0.47        \\ 
		    & TSN  & 13.12     & 17.52     & 8.81      & 22.97      & 10.58         & 6.96            & -1.26        \\ 
		    & SMM  & \B{25.47} & \B{28.84} & \B{16.98} & \B{36.89 } & \B{20.02}     & \B{17.62}      & \B{4.48}     \\ 
		    & Both & \B{25.28}     & 28.00     & \B{16.91}     & 35.93      & 19.08         & 17.32           & \B{4.28}         \\ 
		\hline
		\multirow{4}{*}{BoardGame}    & Base & 35.02     & 35.76     & 29.27     & 39.61      & 19.58         & 18.82           & 10.30        \\ 
		    & TSN  & 34.32     & 35.20     & 28.85     & 39.32      & 18.87         & 17.98           & 9.72         \\ 
		    & SMM  & \B{41.47} & \B{41.37} & \B{35.67} & \B{45.09 } & \B{24.42}     & \B{24.52}       & \B{15.92}    \\ 
		    & Both & 40.40     & 40.54     & 34.16     & 44.16      & 23.45         & 23.32           & 14.26        \\ 
		\bottomrule
	\end{tabular}
	\caption[Evaluation Results: Refactoring-based Obfuscation]{Statistical measures for plagiarism pairs and their differences ($\Delta$) from original pairs for \textbf{refactoring-based obfuscation} (corresponds to \autoref{fig:stage2.5-results}). Higher values indicate better performance. Note that measures are expressed as percentages and their differences as percentage points. Highest values by a margin of 0.25 are marked in bold.}
	\label{tab:diff-refactor}
\end{table}

% ------------------------------------------------------- $
\begin{table}%[h]
\centering
%\small
\setlength{\tabcolsep}{4pt}
\begin{tabular}{lrrrrrrrr}
  \toprule
Dataset & Variant & Pairs & $p$ & $W$ & $\delta$ & $\delta\,Int.$ & $\delta$ 95\% CI & n \\
        \midrule
   \multirow{3}{*}{{PROGpedia-19}}& TSN & P2S & 1 & 11 & -0.036 & Negligible & [-0.33, 0.26] & 27 \\ 
   & SMM & P2S & 6.5e-06 & 325 & 0.567 & Large & [0.28, 0.76] & 27 \\ 
   & Both & P2S & 5e-06 & 350 & 0.510 & Large & [0.21, 0.72] & 27 \\
     \hline
   \multirow{3}{*}{{PROGpedia-56}}& TSN & P2S & 0.31 & 108 & -0.015 & Negligible & [-0.31, 0.28] & 28 \\ 
   & SMM & P2S & 2.2e-05 & 253 & 0.462 & Medium & [0.16, 0.68] & 28 \\ 
   & Both & P2S & 6.5e-06 & 325 & 0.473 & Medium & [0.17, 0.69] & 28 \\ 
   \hline
   \multirow{3}{*}{{TicTacToe}} & TSN & P2S & 0.98 & 143 & -0.020 & Negligible & [-0.24, 0.21] & 50 \\ 
   & SMM & P2S & 5.7e-10 & 1,225 & 0.513 & Large & [0.30, 0.68] & 50 \\ 
   & Both & P2S & 5.7e-10 & 1,225 & 0.484 & Large & [0.27, 0.65] & 50 \\ 
  \hline 
   \multirow{3}{*}{{BoardGame}}& TSN & P2S & 1 & 14 & -0.065 & Negligible & [-0.40, 0.28] & 20 \\ 
   & SMM & P2S & 4.8e-05 & 210 & 0.430 & Medium & [0.06, 0.70] & 20 \\ 
   & Both & P2S & 4.8e-05 & 210 & 0.380 & Medium & [0.01, 0.66] & 20 \\
   \bottomrule
\end{tabular}
\caption[Statistical Tests: Refactoring-based Obfuscation]{One-sided Wilcoxon signed-rank test results for \textbf{refactoring-based obfuscation} regarding the improvement by our defense mechanism compared to baseline (sig. level of $\alpha=0.01$, alternative hypothesis $H1=greater$, test statistic $W$, effect size via Cliff's delta $\delta$, its interpretation $\delta\,Int.$, its 95 percent confidence interval $CI$, and the sample size $n$). For plagiarism-to-source pairs (P2S), low $p$ and high $\delta$ are desirable.} 
\label{tab:to-base-refactor}
\end{table} 
% ------------------------------------------------------- $

\subsubsection*{Baseline}
\autoref{fig:stage2.5-results} shows the substantial impact of refactoring-based obfuscation on the baseline. Median similarity values for plagiarism pairs drop to between 13.90\% (TicTacToe) and 35.02\% (BoardGame), leading to clear overlap with original pairs in all datasets except BoardGame, where overlap is limited to outliers. The reduced effect on BoardGame likely stems from the fact that complex obfuscation techniques are harder to apply broadly, reducing the effectiveness for large programs.
Median similarity \textit{differences} between plagiarism and original pairs (\autoref{tab:diff-refactor}) range from 7.83 (TicTacToe) to 18.82 percentage points (BoardGame), indicating only limited separation. Overall, refactoring proves to be an effective obfuscation method against baseline JPlag.

\subsubsection*{Token Sequence Normalization}
Since refactoring-based obfuscation does not involve statement insertion or reordering, token sequence normalization has little to no effect effect -- an expected outcome, given its design focus.
As shown in \autoref{fig:stage2.5-results}, the results closely resemble the ones for the baseline. In three of four datasets, the median similarity for plagiarism pairs is slightly lower, with reductions ranging from 0.78 (TicTacToe) to 2.73 percentage points (PROGpedia-56) -- a marginal difference.
This trend is also reflected in the similarity \textit{differences} between plagiarism and original pairs (\autoref{tab:diff-refactor}), which range from 6.96 (TicTacToe) to 18.87 percentage points (BoardGame).
Statistical tests (\autoref{tab:to-base-refactor}) confirm that these changes are neither statistically nor practically significant. High p-values and near-zero (even negative) effect sizes indicate negligible impact.
In summary, token sequence normalization provides no measurable resilience against refactoring-based obfuscation, which is consistent with its intended scope.

\subsubsection*{Subsequence Match Merging}
Subsequence match merging leads to clear improvements over the baseline. As shown in \autoref{fig:stage3-results}, median similarity values for plagiarism pairs increase across all datasets, ranging from 25.47\% (TicTacToe) to 42.62\% (PROGpedia-56). Overlap with original pairs is reduced, mostly limited to quartile extremes.
Median similarity \textit{differences} between plagiarism and original pairs (\autoref{tab:diff-refactor}) range from 17.62 (TicTacToe) to 36.02 percentage points (PROGpedia-56), indicating stronger separation.
Statistical tests (\autoref{tab:to-base-refactor}) confirm statistical and practical significance: p-values are low across all datasets, and effect sizes are medium (BoardGame, PROGpedia-56) to large (TicTacToe, PROGpedia-19).
In summary, subsequence match merging provides meaningful resilience against refactoring-based obfuscation, enabling effective detection even when structural changes are introduced through extensive refactorings.

\subsubsection*{Combination of Both}
Combining both defense mechanisms yields strong improvements over the baseline, closely mirroring the results of subsequence match merging alone. As shown in \autoref{fig:stage3-results}, median similarity values for plagiarism pairs range from 25.28\% (TicTacToe) to 41.32\% (PROGpedia-56), with limited overlap confined to quartile extremes.
Median similarity \textit{differences} (\autoref{tab:diff-refactor}) span from 17.32 (TicTacToe) to 34.58 percentage points (PROGpedia-56), showing solid improvement. Results are slightly lower than those of subsequence match merging alone, except for PROGpedia-56, where the combination performs marginally better.
Statistical tests (\autoref{tab:to-base-refactor}) confirm both statistical and practical significance: all p-values are low, with effect sizes ranging from medium (BoardGame, PROGpedia-56) to large (TicTacToe, PROGpedia-19).
In summary, the combined defenses offer significant resilience against refactoring-based obfuscation across all datasets, though the added benefit over subsequence match merging alone is limited.

\summaryBox{3}{The defense mechanisms significantly increase the resilience against semantic preserving refactoring-based obfuscation attacks. The median similarity differences increase, depending on the dataset, up to 22 percentage points, thus strongly improving the separation between plagiarized and original programs.}

\subsection{GPT-4-based Obfuscation}\label{sec:eval-gptobf}
\begin{figure}
\centering
\includegraphics[width=\linewidth, trim=0 5pt 0 0, clip]{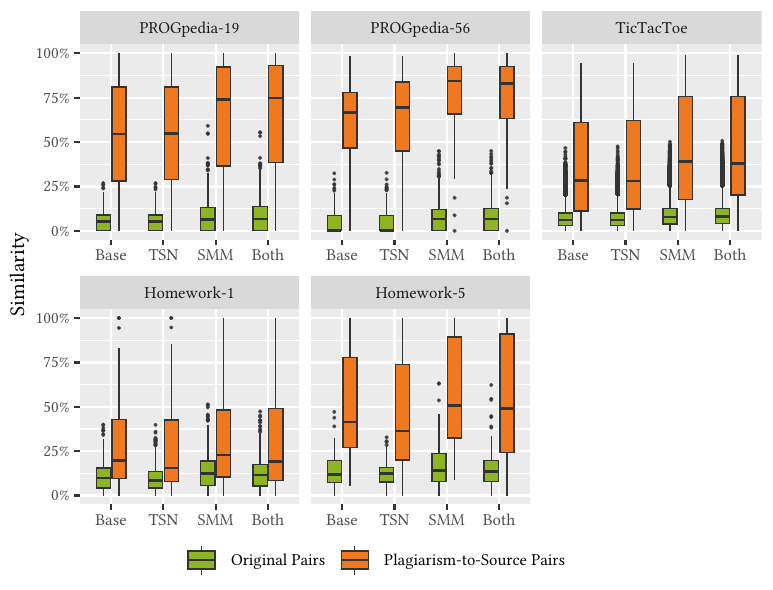}
\caption[Evaluation Results: AI-based Obfuscation]{Similarity scores for original program pairs and \textbf{AI-based plagiarism} pairs (obfuscation with 15 varying GPT-4 prompts). Ideally, plagiarism pairs exhibit high similarity, while original pairs should exhibit low similarity. }
\label{fig:stage3-results}
\end{figure}

\autoref{fig:stage3-results} presents the results for GPT-4-based obfuscation, with corresponding statistical measures in \autoref{tab:diff-gpt-obf}. We used 15 distinct prompts instructing GPT-4 to alter program code while preserving its functionality. However, as this process lacks formal guarantees, the resulting obfuscation is considered semantic-agnostic.
The BoardGame dataset was excluded due to its use in a final exam and the associated privacy concerns with sending data to OpenAI servers.

\begin{table}
	\centering
    \setlength{\tabcolsep}{3pt}
	\begin{tabular}{lrrrrrrrr}
		\toprule
		Dataset                       & Variant & Median    & Mean      & $Q_1$     & $Q_3$     & $\Delta$ Mean & $\Delta$ Median & $\Delta$ IQR \\ 
		\midrule
		\multirow{4}{*}{PROGpedia-19} & Base     & 54.59     & 54.12     & 28.06     & 81.04     & 48.39         & 49.53           & 19.15        \\ 
		                               & TSN      & 54.74     & 54.84     & 28.82     & 80.95     & 49.08         & 49.61           & 19.88        \\ 
		                               & SMM      & 73.95     & \B{63.46} & 36.51     & 92.34     & \B{54.31}     & 67.59           & 23.31        \\ 
		                               & Both     & \B{74.79} & \B{63.60} & \B{38.32} & \B{92.98} & \B{54.38}    & \B{68.22}       & \B{24.44}    \\ 
		\hline
		\multirow{4}{*}{PROGpedia-56} & Base     & 66.67     & 61.76     & 46.67     & 77.96     & 56.87         & 66.67           & 37.91        \\ 
		                               & TSN      & 69.40     & 63.92     & 45.05     & 83.92     & 59.40         & 69.40           & 36.41        \\ 
		                               & SMM      & \B{84.43} & \B{75.37} & \B{65.76} & \B{92.60} & \B{67.11}     & \B{77.84}       & \B{53.56}    \\ 
		                               & Both     & 83.07     & 75.10     & 63.11     & \B{92.59} & \B{66.98}     & 76.32           & 50.50        \\ 
		\hline
		\multirow{4}{*}{TicTacToe}    & Base     & 28.20     & 35.60     & 11.25     & 61.05     & 28.75         & 22.13           & 1.28         \\ 
		                               & TSN      & 27.98     & 37.50     & 12.32     & 62.26     & 30.56         & 21.83           & 2.25         \\ 
		                               & SMM      & \B{39.02} & 44.96     & 17.68     & \B{75.65} & 36.13         & \B{31.18}       & 5.16         \\ 
		                               & Both     & 38.07     & \B{45.53} & \B{20.11} & \B{75.70} & \B{36.61}     & 30.13           & \B{7.48}     \\ 
		\hline
		\multirow{4}{*}{Homework-1}   & Base     & 19.74     & 27.71     & 9.70      & 42.81     & 17.32         & 9.86            & \B{-5.90}    \\ 
		                               & TSN      & 15.47     & 26.44     & 7.88      & 42.50     & 17.07         & 6.83            & \B{-5.77}    \\ 
		                               & SMM      & \B{22.90} & \B{32.69} & \B{10.49} & 48.27     & \B{19.65}     & \B{10.32}       & -9.01        \\ 
		                               & Both     & 19.09     & 31.42     & 8.35      & \B{48.96} & \B{19.42}     & 7.47            & -9.08        \\ 
		\hline
		\multirow{4}{*}{Homework-5}   & Base     & 41.33     & 50.37     & 27.02     & 77.79     & 36.97         & 29.34           & 7.34         \\ 
		                               & TSN      & 36.26     & 46.13     & 20.03     & 73.68     & 34.48         & 23.98           & 4.19         \\ 
		                               & SMM      & \B{50.68} & \B{57.54} & \B{32.45} & 89.24     & \B{41.62}     & \B{36.70}       & \B{8.79}     \\ 
		                               & Both     & 49.06     & 53.90     & 24.40     & \B{90.91} & 39.31         & 35.71           & 4.82         \\ 
		\bottomrule  
	\end{tabular}
	\caption[Evaluation Results: AI-based Obfuscation]{Statistical measures for plagiarism pairs and their differences ($\Delta$) from original pairs for \textbf{AI-based obfuscation} (corresponds to \autoref{fig:stage3-results}). Higher values indicate better performance. Note that measures are expressed as percentages and their differences as percentage points. Highest values by a margin of 0.25 are marked in bold.}
	\label{tab:diff-gpt-obf}
\end{table}

% ------------------------------------------------------- $
\begin{table}%[h]
\centering
%\small
\setlength{\tabcolsep}{4pt}
\begin{tabular}{lrrrrrrrr}
  \toprule
Dataset & Variant & Pairs & $p$ & $W$ & $\delta$ & $\delta\,Int.$ & $\delta$ 95\% CI & n \\ 
  \midrule 
   \multirow{3}{*}{{PROGpedia-19}} & TSN & P2S & 0.0027 & 893 & 0.019 & Negligible & [-0.17, 0.20] & 74 \\ 
   & SMM & P2S & $<$ 1e-10 & 1,485 & 0.210 & Small & [0.02, 0.38] & 74 \\ 
   & Both & P2S & $<$ 1e-10 & 2,216 & 0.209 & Small & [0.02, 0.38] & 74 \\ 
      \hline 
   \multirow{3}{*}{{PROGpedia-56}} & TSN & P2S & 0.025 & 971 & 0.077 & Negligible & [-0.11, 0.26] & 75 \\ 
   & SMM & P2S & $<$ 1e-10 & 1,540 & 0.371 & Medium & [0.19, 0.53] & 75 \\ 
   & Both & P2S & $<$ 1e-10 & 2,136 & 0.360 & Medium & [0.18, 0.52] & 75 \\ 
      \hline
   \multirow{3}{*}{{TicTacToe}} & TSN & P2S & 5.2e-08 & 1,107 & 0.042 & Negligible & [-0.14, 0.22] & 75 \\ 
   & SMM & P2S & $<$ 1e-10 & 1,770 & 0.194 & Small & [0.01, 0.37] & 75 \\ 
   & Both & P2S & $<$ 1e-10 & 2,342 & 0.210 & Small & [0.03, 0.38] & 75 \\ 
      \hline
   \multirow{3}{*}{{Homework-1}} & TSN & P2S & 0.89 & 913 & -0.046 & Negligible & [-0.23, 0.14] & 74 \\ 
   & SMM & P2S & 2e-06 & 406 & 0.075 & Negligible & [-0.11, 0.26] & 74 \\ 
   & Both & P2S & 0.003 & 1,536 & 0.024 & Negligible & [-0.16, 0.21] & 74 \\ 
      \hline
   \multirow{3}{*}{{Homework-5}}& TSN & P2S & 1 & 229 & -0.115 & Negligible & [-0.29, 0.07] & 75 \\ 
   & SMM & P2S & 2.7e-09 & 1,035 & 0.153 & Small & [-0.03, 0.33] & 75 \\ 
   & Both & P2S & 0.11 & 1,260 & 0.047 & Negligible & [-0.14, 0.23] & 75 \\ 
   \bottomrule
\end{tabular}
\caption[Statistical Tests: AI-based Obfuscation]{One-sided Wilcoxon signed-rank test results for \textbf{AI-based obfuscation} regarding the improvement by of our defense mechanism compared to baseline (sig. level of $\alpha=0.01$, alternative hypothesis $H1=greater$, test statistic $W$, effect size via Cliff's delta $\delta$, its interpretation $\delta\,Int.$, its 95 percent confidence interval $CI$, and the sample size $n$). For plagiarism-to-source pairs (P2S), low $p$ and high $\delta$ are desirable.} 
\label{tab:to-base-gpt-obf}
\end{table}
 
% ------------------------------------------------------- $

\subsubsection*{Baseline}
\autoref{fig:stage3-results} shows that GPT-4-based obfuscation, using 15 behavior-preserving prompts, can be effective against JPlag. Median similarity values for plagiarism pairs drop to between 19.74\% (Homework-5) and 66.67\% (PROGpedia-56), which is notably a higher range than seen with other obfuscation methods.
Overlap with original pairs varies across datasets. Homework-1 shows substantial overlap, including interquartile ranges, while overlap is less pronounced for PROGpedia datasets. As shown in \autoref{tab:diff-gpt-obf}, median similarity \textit{differences} range from 9.86 (Homework-1) to 66.67 percentage points (PROGpedia-56), indicating limited separation -- particularly for Homework and TicTacToe.
Interestingly, the variability in attack effectiveness across prompts is similar to that across datasets.
We observe that the dataset itself -- likely due to the underlying assignment and domain -- has a greater influence on obfuscation effectiveness.
Thus, while GPT-based obfuscation is effective, its reliability is lower than that of algorithmic methods due to significant variation in performance across datasets and prompts.

\subsubsection*{Token Sequence Normalization}
Since GPT-4-based obfuscation involves diverse modifications beyond statement insertion or reordering, token sequence normalization has a limited impact. As shown in \autoref{fig:stage3-results}, results are similar to the baseline across all five datasets.
Median similarity values for plagiarism pairs vary slightly, ranging from $-5.07$ (Homework-5) to $+2.37$ percentage points (PROGpedia-56). Corresponding median similarity \textit{differences} with original pairs (\autoref{tab:diff-gpt-obf}) range from 6.83 (Homework-1) to 69.40 percentage points (PROGpedia-56), aligning closely with baseline values -- slightly better for PROGpedia, slightly worse for the Homework datasets, possibly reflecting language-specific differences in GPT-4’s output.
Statistical tests (\autoref{tab:to-base-gpt-obf}) show some statistical significance (PROGpedia-19, TicTacToe) but no practical significance. Effect sizes are negligible across all datasets, with negative values for the Homework sets.
In summary, token sequence normalization offers no meaningful resilience against GPT-4-based obfuscation, though it also introduces no adverse effects.

\subsubsection*{Subsequence Match Merging}
Subsequence match merging significantly improves results over the baseline. As shown in \autoref{fig:stage3-results}, median similarity values for plagiarism pairs increase across all datasets, ranging from 22.90\% (Homework-1) to 84.43\% (PROGpedia-56). Overlap with original pairs is reduced, primarily limited to quartile extremes.
Median similarity \textit{differences} (\autoref{tab:diff-gpt-obf}) range from 10.32 (Homework-1) to 77.84 percentage points (PROGpedia-56), indicating substantial separation. This confirms that subsequence match merging provides a solid improvement over the baseline.
Statistical tests (\autoref{tab:to-base-gpt-obf}) show statistically significant improvements across all datasets. Practical significance is achieved in all but Homework-1, where the effect size remains negligible. Note that the high variance in plagiarism pair similarities affects the effect size measure~\cite{Grissom2012}. In contrast, the remaining datasets show small to medium effect sizes, indicating practical significance.
In sum, subsequence match merging offers robust resilience against GPT-4-based obfuscation despite its semantic-agnostic nature and variability across datasets and prompts.

\subsubsection*{Combination of Both}
Combining both defense mechanisms results in strong improvements over the baseline, largely mirroring the effect of subsequence match merging alone. As shown in \autoref{fig:stage3-results}, median similarity values for plagiarism pairs range from 19.09\% (Homework-1) to 83.07\% (PROGpedia-56), with reduced overlap mostly confined to quartile boundaries.
Median similarity \textit{differences} (\autoref{tab:diff-gpt-obf}) range from 7.47 (Homework-1) to 76.32 percentage points (PROGpedia-56), showing consistent improvement over the baseline. The effect is slightly weaker than with subsequence match merging alone, except for PROGpedia-19, where the combination performs marginally better.
Statistical tests (\autoref{tab:to-base-gpt-obf}) confirm statistical significance in all datasets except Homework-5 ($p=0.11$). Practical significance is observed for all but the Homework datasets, which exhibit small effect sizes. These datasets appear more vulnerable to AI-based obfuscation, possibly due to the small size of these programs or due to the semantic-agnostic nature of the transformation, which may alter the behavior of programs.
For the remaining four Java datasets, effect sizes are small to medium, indicating practical significance. As with other AI-based attacks, high variance in similarity scores due to prompt diversity reduces measured effect sizes~\cite{Grissom2012}.
In summary, the combined defenses offer significant resilience against AI-based obfuscation for Java datasets, though results are more limited for C++ programs. Yet, the defense mechanisms improve detection despite the potentially disruptive nature of AI-based obfuscation.

\summaryBox{4}{The defense mechanisms significantly increase the resilience against semantic agnostic AI-based obfuscation attacks. The median similarity differences increase, depending on the dataset, up to 19 percentage points, thus improving the separation between plagiarized and original programs, albeit to a lesser degree than other attack types.}

\subsection{GPT-4-generated Programs}\label{sec:eval-gptgen}
\begin{figure}
\centering
\includegraphics[width=\linewidth, trim=0 5pt 0 0, clip]{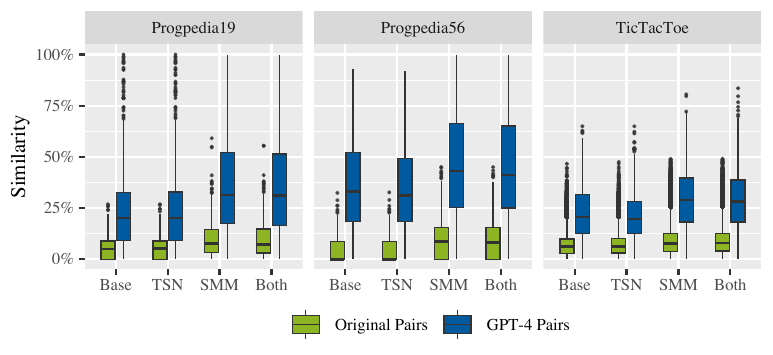}
\caption[Evaluation Results: AI-based Generation]{Similarity scores for original (human) program pairs and pairs of \textbf{AI-generated programs} (based on GPT-4 and the assignment description). Ideally, generated pairs exhibit high similarity, while original pairs should exhibit low similarity.}
\label{fig:stage4-results}
\end{figure}

\autoref{fig:stage4-results} presents results for programs generated by GPT-4 based on assignment descriptions, with corresponding statistical measures in \autoref{tab:diff-gpt-gen}. Unlike previous stages, this evaluation does not involve obfuscation, as the generated programs are not derived from human-written ones. Instead, we compare the similarity among GPT-4-generated programs to that of unrelated human submissions.
While the defense mechanisms are not designed for this setting, they improve the distinction between AI-generated and unrelated human programs. Such a capability can help detect AI-generated submissions if multiple students use the same language model.
As with the last stage, BoardGame was excluded for privacy reasons.

\begin{table}
	\centering
    \setlength{\tabcolsep}{3pt}
\begin{tabular}{lrrrrrrrrr}
	\toprule
	Dataset                       & Variant & Median    & Mean      & $Q_1$      & $Q_3$     & $\Delta$ Mean & $\Delta$ Median & $\Delta$ IQR \\ 
	\midrule
	\multirow{4}{*}{PROGpedia-19} & Base    & 20.10     & 22.90     & 9.09       & 32.77     & 17.17         & 15.04           & 0.18         \\ 
	                               & TSN     & 20.20     & 23.20     & 9.18       & 32.93     & 17.44         & 15.07           & 0.24         \\ 
	                               & SMM     & \B{31.33} & \B{36.66} & \B{17.51}  & \B{52.14} & \B{26.52}     & 23.58           & \B{2.92}     \\ 
	                               & Both    & \B{31.25} & 36.33     & 16.49      & 51.35     & \B{26.38}     & \B{23.96}       & 1.73         \\
	\hline
	\multirow{4}{*}{PROGpedia-56} & Base    & 33.13     & 35.81     & 18.53      & 52.04     & 30.92         & 33.13           & 9.77         \\ 
	                               & TSN     & 31.25     & 34.58     & 18.45      & 49.11     & 30.06         & 31.25           & 9.81         \\ 
	                               & SMM     & \B{43.11} & \B{45.50} & \B{25.30}  & \B{66.20} & \B{35.31}     & \B{34.47}       & \B{9.85}     \\ 
	                               & Both    & 41.18     & 44.99     & \B{25.10}  & 65.13     & \B{35.24}     & 32.91           & 9.59         \\
	\hline
	\multirow{4}{*}{TicTacToe}    & Base    & 20.63     & 22.53     & 12.53      & 31.51     & 15.67         & 14.57           & 2.57         \\ 
	                               & TSN     & 19.62     & 20.72     & 12.47      & 28.25     & 13.78         & 13.46           & 2.40         \\ 
	                               & SMM     & \B{28.94} & \B{29.60} & \B{18.27}  & \B{39.72} & \B{20.77}     & \B{21.10}       & \B{5.75}     \\ 
	                               & Both    & 28.18     & 28.98     & \B{18.18}  & 38.65     & 20.06         & 20.24           & \B{5.55}         \\ 
	\bottomrule
\end{tabular}
	\caption[Evaluation Results: AI-based Generation]{Statistical measures for plagiarism pairs and their differences ($\Delta$) from original pairs for \textbf{AI-based generation} (corresponds to \autoref{fig:stage4-results}). Higher values indicate better performance. Note that measures are expressed as percentages and their differences as percentage points. Highest values by a margin of 0.25 are marked in bold.}
	\label{tab:diff-gpt-gen}
\end{table}

% ------------------------------------------------------- $
\begin{table}%[h]
	\centering
	%\small
    \setlength{\tabcolsep}{3pt}
	\begin{tabular}{lrrrrrrrr}
		\toprule
		Dataset                      & Variant & Pairs & $p$       & $W$     & $\delta$ & $\delta\,Int.$ & $\delta$ 95\% CI & n     \\ 
		\midrule
		\multirow{3}{*}{PROGpedia-19}  & TSN     & FG    & 1.8e-11   & 4,937   & 0.007    & Negligible     & [-0.04,  0.05]    & 1,225 \\ 
		                               & SMM     & FG    & $<$ 1e-10 & 445,096 & 0.321    & Small          & [0.28,  0.36]     & 1,225 \\ 
		                               & Both    & FG    & $<$ 1e-10 & 407,953 & 0.310    & Small          & [0.27,  0.35]     & 1,225 \\ 
		\hline
		\multirow{3}{*}{PROGpedia-56}  & TSN     & FG    & 1         & 175,838 & -0.033   & Negligible     & [-0.08,  0.01]    & 1,225 \\ 
		                               & SMM     & FG    & $<$ 1e-10 & 475,800 & 0.219    & Small          & [0.17,  0.26]     & 1,225 \\ 
		                               & Both    & FG    & $<$ 1e-10 & 636,011 & 0.207    & Small          & [0.16,  0.25]     & 1,225 \\ 
		\hline
		\multirow{3}{*}{{TicTacToe}}  & TSN     & FG    & 1         & 120,687 & -0.072   & Negligible     & [-0.12, -0.03]   & 1,225 \\ 
		                               & SMM     & FG    & $<$ 1e-10 & 334,971 & 0.271    & Small          & [0.23,  0.31]     & 1,225 \\ 
		                               & Both    & FG    & $<$ 1e-10 & 416,672 & 0.256    & Small          & [0.21,  0.30]     & 1,225 \\ 
		\bottomrule
	\end{tabular}
	\caption[Statistical Tests: AI-based Generation]{One-sided Wilcoxon signed-rank test results for \textbf{AI-based generation} regarding the improvement by our defense mechanism compared to baseline (sig. level of $\alpha=0.05$, alternative hypothesis $H1=greater$, test statistic $W$, effect size via Cliff's delta $\delta$, its interpretation $\delta\,Int.$, its 95 percent confidence interval $CI$, and the sample size $n$). For Fully-Generated Pairs (FG), low $p$ and high $\delta$ are desirable.} 
	\label{tab:to-base-gpt-gen}
\end{table}

% ------------------------------------------------------- $

\subsubsection*{Baseline}
\autoref{fig:stage4-results} shows that GPT-4-generated programs, though created from the same assignment prompt, exhibit relatively low similarity: the median similarities among generated pairs are between 20.10\% (PROGpedia-19) and 31.33\% (PROGpedia-56). This reflects the inherent indeterminism of generative AI.
In contrast, however, unrelated human-made programs for the same task have even lower similarities, with median values between 0.00\% (PROGpedia-56) and 6.06\% (TicTacToe). Thus, even with the baseline, GPT-4-generated programs are significantly more similar to each other than human submissions.

However, some overlap remains. As shown in \autoref{tab:diff-gpt-gen}, the median similarity \textit{differences} between AI-generated and human program pairs are 14.57 (TicTacToe), 15.04 (PROGpedia-19) and 33.13 (PROGpedia-56) percentage points, which is comparable to differences observed for refactoring- and alteration-based obfuscation, but notably higher than for insertion-based attacks, which showed a median difference slightly below zero ($-0.78$).
In summary, while generated programs are more alike than human ones, the limited separation leaves room for evasion -- highlighting the need for improved detection mechanisms.

\subsubsection*{Token Sequence Normalization}
As token sequence normalization targets statement insertion and reordering, it is not expected to meaningfully affect AI-generated programs, which typically lack dead code and are less impacted by statement order.
As shown in \autoref{fig:stage4-results}, results align with expectations: the median similarities among generated pairs match the baseline. Similarly, the median similarity \textit{difference} with unrelated human programs are at similar values as the baseline (\autoref{tab:diff-gpt-gen}).
Statistical tests (\autoref{tab:to-base-gpt-gen}) confirm that the variations are both statistically and practically insignificant. With a p-value of 1 and a near-zero effect size, no meaningful improvement is observed.
In summary, token sequence normalization does not enhance the detection of AI-generated programs but also introduces no significant drawbacks.

\subsubsection*{Subsequence Match Merging}
Subsequence match merging yields a clear improvement over the baseline. As shown in \autoref{fig:stage4-results}, the median similarity among generated pairs increases by more than 8.31 (TicTacToe) to 11.23 (PROGpedia-19) percentage points, reducing overlap with human programs -- now mostly limited to the upper quartile.
This improvement is reflected in the median similarity \textit{difference} between AI-generated and human program pairs, which rises to between 21.10 (TicTacToe) and 34.47 (PROGpedia-56) percentage points (\autoref{tab:diff-gpt-gen}).
Statistical tests (\autoref{tab:to-base-gpt-gen}) confirm both statistical and practical significance: p-values are low, and the effect size, though small, is meaningful in practice.
In summary, subsequence match merging significantly enhances the detection of AI-generated programs -- despite not being designed for this purpose -- highlighting its versatility as a defense mechanism.

\subsubsection*{Combination of Both}
Combining both defense mechanisms results in a strong improvement over the baseline, closely mirroring the effect of subsequence match merging alone. As shown in \autoref{fig:stage4-results}, the median similarity among generated pairs increases by between 7.55 (TicTacToe) and 11.15 (PROGpedia-19) percentage points, reducing overlap with human submissions -- primarily in the upper quartile.
The median similarity \textit{difference} between AI-generated and human program pairs rises to between 20.24 (TicTacToe) and 34.47 (PROGpedia-56) percentage points (\autoref{tab:diff-gpt-gen}). 
Statistical tests (\autoref{tab:to-base-gpt-gen}) confirm both statistical and practical significance, with low p-values and a small but meaningful effect size.
In summary, the combination of both defenses significantly improves the detection of AI-generated programs, performing nearly identically to subsequence match merging alone and introducing no observable drawbacks.

\summaryBox{5}{The defense mechanisms, while not designed for this purpose, significantly increase the detection rate of AI-generated programs. The median similarity difference to human programs increases by up to 8.92 percentage points, thus improving the separation between plagiarized and original programs moderately but yet significantly.}

\subsection{Threshold-based Obfuscation}\label{sec:eval-mossad}
In the previous sections, we evaluated each defense variant using identical plagiarism instances, enabling direct comparison. However, threshold-based obfuscation (e.g., \mossad \cite{DevoreMcDonald2020}) dynamically adapts its transformations based on the output of a plagiarism detector, as it obfuscates until the similarity to the original falls below a target threshold. In \autoref{sec:eval-insert}, we configured it to use baseline JPlag. This raises the question of how the defenses affect obfuscation when enabled during the threshold-based obfuscation process itself.
To investigate this, we selected ten random programs from each Homework dataset and ran \mossad with all defense variants (\mossad only supports C/C++ programs). Note that these programs are relatively small (105 LOC on average in Homework-1 and 282 LOC in Homework-5). Experiments were run on a high-performance system (AMD Ryzen 7 7700, 16GB RAM, Arch Linux), representing a realistic upper bound for student hardware.
%%%
For all programs and variants, \mossad is configured to terminate when reaching the 25 percent similarity to the original.
We observed a notable increase in obfuscation time and program size when using \mossad to evade the defense mechanisms. 

\begin{samepage}
\begin{figure}
\centering
\includegraphics[width=\linewidth, trim=0 12pt 0 0, clip]{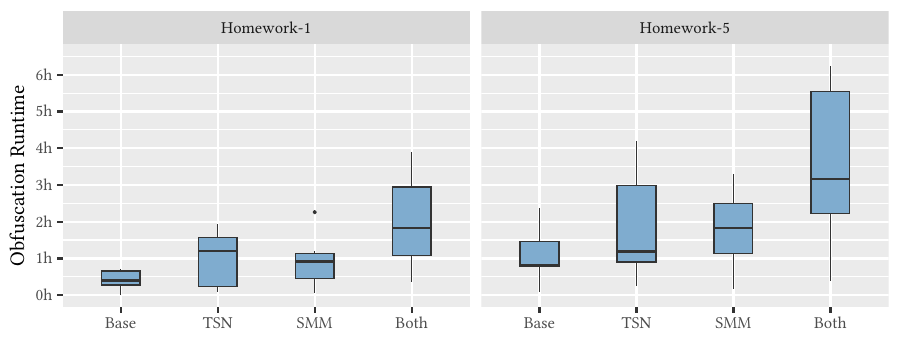}
\caption[Evaluation Results: Obfuscation Duration]{Required obfuscation duration per program for \mossad to reach an obfuscation threshold of 25 percent (for programs with original sizes of \textasciitilde105 LOC for Hw.-1 and \textasciitilde123 LOC for Hw.-5).}
\label{fig:stage5-result-time}
\vspace{10pt}
\centering
\includegraphics[width=\linewidth, trim=0 12pt 0 0, clip]{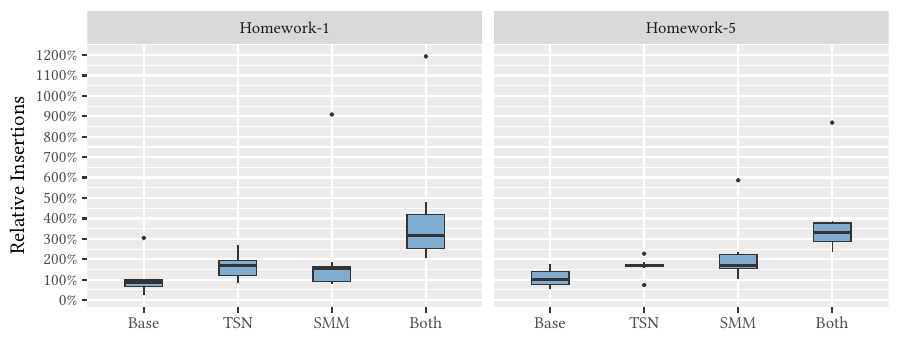}
\caption[Evaluation Results: Inserted Statements]{Required relative insertion of statements for \mossad to reach a 25 percent obfuscation threshold (relative insertions compared to the original program size to normalize for program size).}
\label{fig:stage5-result-loc}
\end{figure}
\end{samepage}

\autoref{fig:stage5-result-time} shows a substantial increase in runtime when defenses are active. For the baseline, median obfuscation times were 24 minutes (Homework-1) and 48 minutes (Homework-5).  
With token sequence normalization (TSN), runtime increased to 72 minutes across both datasets.  
With subsequence match merging (SMM), runtimes rose to 55 minutes (Homework-1) and 110 minutes (Homework-5).  
Combining both defenses further increased runtimes to 110 minutes and 190 minutes, respectively, with observed maxima reaching 234 and 375 minutes.
These results show that the defenses impose significant computational overhead even on small programs, raising the effort required to successfully obfuscate plagiarism.

\autoref{fig:stage5-result-loc} shows the relative growth in program size. For the baseline, the median size increased by 87\% (Homework-1) and 101\% (Homework-5), nearly doubling the original code.  
With TSN, insertions rose to 169\%, while SMM required 156\% (Homework-1) and 169\% (Homework-5).  
When both defenses were enabled, median increases reached 317\% and 331\%, respectively. Maximum observed increases were 1193\% and 869\%, and all obfuscated programs exceeded 200\% of the original size.  
Such extreme growth makes obfuscated code highly conspicuous and easily identifiable through token count, manual inspection, or outlier detection.
Consequently, it makes threshold-based obfuscation apparent, thus turning it into an ineffective obfuscation strategy.

Despite the high-performance system and small input programs, obfuscating just 20 programs across all variants took over 125 hours. Larger programs or slower systems would require days per attempt.  
Importantly, we configured \mossad to stop at a 25\% similarity threshold (to restrict then overall computation time), while unrelated program pairs typically fall around 10–15\%. Achieving lower similarity -- and thus avoiding detection entirely -- would require even more aggressive obfuscation.
Overall, our contributions substantially enhance obfuscation resilience, making threshold-based obfuscation highly time-consuming and resulting in plagiarized solutions that are exceptionally conspicuous due to their size. These factors collectively act as strong deterrents against obfuscation-based plagiarism, making the obfuscation efforts more tedious than completing the actual assignment.

\summaryBox{5}{The defense mechanisms strongly increase the computational cost for threshold-based plagiarism, thus resulting in an obfuscation time of up to 6 hours per program and up to 1300 percent increase in program size, making threshold-based plagiarism more tedious and easily detectable.}

\section{Threats to Validity}\label{sec:six}
We now discuss how we address threats to the validity of our evaluation, following the guidelines outlined by \citet{Wohlin2012} and \citet{runeson2008}. 

\subsection*{Internal Validity} 
    \textit{Baseline Consistency:} For internal validity, we used JPlag as a baseline but also implemented the defense mechanism for JPlag, ensuring that all other conditions remained constant when comparing the defense mechanism with each other or with the baseline. 
    \textit{Handling of invalid programs:} Some public datasets contain invalid or incomplete programs (e.g., programs that do not compile), which could lead to inaccurate results if not properly handled. We addressed this by preprocessing the datasets and removing programs that do not compile.
    \textit{Validity of the Labeling:}
    Public datasets often contain incomplete or biased plagiarism labels. This issue does not affect our evaluation, as all plagiarism instances are generated through controlled automated obfuscation.
    As a preprocessing step, we carefully filtered out instances of human plagiarism based on the labels, analyzed them with JPlag, and performed human inspections.

\subsection*{External Validity}
    \textit{Generalizability across datasets:}
    Our evaluation uses real-world student submissions from diverse university courses, covering two programming languages and varying assignment sizes. This reflects typical software plagiarism detection scenarios and supports a representative, generalizable assessment~\cite{Novak2019}.
    \textit{Generalizability of obfuscation attacks:}
    Limiting the evaluation to only a few types of obfuscation attacks could hinder the applicability of our results to broader contexts. To enhance external validity and thus ensure that our findings are generalizable, we included a diverse set of real-world obfuscation techniques.
    \textit{Influence of Prompt Quality:} To address the impact of prompt choice for AI-based obfuscation, we performed systematic "\textit{prompt-engineering}" prior to the evaluation. We then evaluated with 15 suitable prompts. We generated multiple plagiarism instances for each prompt, which we repeated for multiple datasets. While the impact of the prompt varies, the variation is not strong enough to obscure the overall trend, supporting the generalizability of our results.

\subsection*{Construct Validity}
    \textit{Evaluation Methodology Alignment:} To enhance construct validity, we aligned our evaluation methodology with those from established and related research works. 
    Moreover, we employ an approach-independent ground truth, and use established similarity metrics.
    \textit{Underlying Research Object:} Our measurements align directly with the research objective of evaluating detector resilience against automated obfuscation. We use similarity scores from the detectors as primary measurements and assess obfuscation using real-world tools like \mossad and GPT-4.
    \textit{Choice of Baseline:} The baseline selection might affect the comparison and outcomes.
    We selected JPlag as a baseline, as other widely used tools are either not applicable to all datasets, closed-source, or provide restricted results.
    JPlag is widely recognized as a state-of-the-art tool~\cite{Aniceto2021, Novak2019}, ensuring that the comparison is relevant and accurate. It operates similarly to other widely used tools by employing standard similarity metrics.

\subsection*{Reliability}
To ensure reliability, we provide a comprehensive reproduction package for our evaluation~\cite{replication-package}.
    \textit{Use of Internal Datasets:} Using internal datasets can hinder reproducibility. To enhance reliability, we used both public and internal datasets, balancing generalizability with the need for open data where possible. We discussed any preprocessing steps and the employed obfuscation attacks for all datasets.
    For the internal datasets (TicTacToe and BoardGame), we provide raw results and metadata in our replication package. 
    \textit{Publishing of Obfuscation Attacks:} The obfuscation attacks utilized in our study can be considered malware, which restricts our ability to provide access to these tools. The exception is GPT-4~\cite{gpt4}, which is publicly available; however, we do not provide a detailed, step-by-step guide on exploiting it for plagiarism detection. While omitting these artifacts or details may hinder reproducibility, balancing this limitation with ethical considerations and the responsibility regarding potential misuse. 

\section{Discussion}\label{sec:eval-discussion}\label{sec:seven}
In the following, we discuss the interpretation of the evaluation results and highlight key takeaways for software plagiarism detection.
Our evaluation highlights the effectiveness of automated obfuscation techniques against plagiarism detectors \textit{without} defense mechanisms. Insertion-based obfuscation proves especially effective, fully concealing plagiarism by adding semantically irrelevant code. Refactoring-based obfuscation also poses a substantial challenge, as structural changes that preserve behavior significantly reduce similarity, limiting the detector’s ability to identify plagiarism instances.

AI-based obfuscation introduces significant variability, with its effectiveness depending more on dataset characteristics than on prompt or language. While powerful, its reliability is lower than that of algorithmic methods. However, due to the rapid advancements of generative AI, AI-based poses a growing challenge to detection systems. Currently, AI-based program generation is only effective for smaller programs (below 300 to 400 LOC).
Our results show that AI-generated programs exhibit increased similarity to each other compared to human-written programs, aiding detection when multiple students use the same model. Although not designed for this setting, subsequence match merging improves the separation of AI-generated from human-written programs. 

\subsection{On Providing Broad Obfuscation Resilience}

Our evaluation shows that our approach improves obfuscation resilience for all employed attacks and datasets. As expected, the \textit{degree} of those improvements depends on the type of obfuscation attack. Nevertheless, when employing the defense mechanisms, we demonstrate that the provided resilience is not limited to a specific obfuscation attack. 
We demonstrated effectiveness against a wide-range of obfuscation attacks, including both algorithmic and AI-based attacks, encompassing semantic-preserving and semantic-agnostic obfuscation.
Furthermore, we evaluate datasets across different programming languages, in addition to diverse assignment types and sizes, thus demonstrating its adaptability.
In total, we use six datasets in combination with five distinct obfuscation attack types. Moreover, each attack type involves various modifications. For example, refactoring-based obfuscation includes multiple transformation types, while AI-based obfuscation involves 15 varying prompts to generate diverse plagiarism instances.

Our evaluation showed that our contributions provide broad resilience against automated obfuscation attacks on programming assignments by systematically covering these different categories of obfuscation attacks. The smallest improvement was observed for AI-based obfuscation, which is expected, given that this is a semantic-agnostic attack using highly challenging prompts, including partial implementations. Detecting partial implementations is particularly difficult for plagiarism detectors as they must carefully balance between detecting re-implementation and avoiding false positives.
On the other hand, the strongest improvement was observed for structural attacks, which is a significant result. Structural attacks are among the easiest to automate, even with traditional methods, and they tend to consistently affect plagiarism detectors. Thus, improving resilience in this area is crucial for the effectiveness of detection tools.

\subsection{Outliers and Remaining Overlap}

Except for insertion-based obfuscation (see \autoref{fig:stage1-results}), where the defense mechanisms completely eliminate \textit{any} overlap between plagiarism instances and unrelated programs, the evaluation results demonstrate minor overlap. This raises an important question regarding the expectations one should have concerning the quality of plagiarism detection.
In practical terms, some overlap among outliers is not a significant concern. It is essential to recognize that no plagiarism detection tool is perfect. Thus, educators must accept that human inspection is always the final step in plagiarism detection and that no one should solely rely on the results of an automated tool without first verifying the flagged candidates themselves. 

Furthermore, it is crucial to note that plagiarism detectors compare pairs of programs, and thus, a single program might be included in multiple comparisons. This means that detecting every plagiarism pair is not necessary to identify all students involved in plagiarism. In practice, educators would be presented with a ranked list of suspicious pairs, including unrelated and plagiarism pairs.
For example, in our evaluation of GPT-4-generated programs, it is not necessary to identify all 1,225 pairs of AI-generated programs to detect each of the 50 generated programs at least once.
Notably, when both of the defense mechanisms are enabled (\textit{Both} in \autoref{fig:stage4-results}), only the first 158 pairs (which is the top 0.07 percent out of all 220,780 analyzed pairs) need to be inspected to successfully identify 90 percent of the AI-generated programs at least once. To detect all 50 AI-generated programs, the first 711 pairs need to be checked, which is the top 0.3 percent of all pairs. This underscores that a slight overlap between the pairs of unrelated programs and the plagiarism pairs is not a cause for concern.

Ultimately, it is important to emphasize that no plagiarism detection tool can provide 100 percent certainty. Therefore, human inspection and informed decision-making are essential in ensuring fair and accurate misconduct investigations. Educators must \textit{always} engage in thoughtful analysis of the results generated by these tools to discern genuine cases of plagiarism from false positives effectively.

\subsection{AI-based Plagiarism}\label{sec:discussion-ai}

AI-based attacks~\cite{Biderman2022}, particularly those utilizing generative AI, present a growing concern for plagiarism detection.
We discussed two possible scenarios when employing generative AI to cheat on programming assignments. \textit{Automatic obfuscation} of an existing solution and \textit{fully generating} solutions from the assignment description.
Based on our evaluation results, automatic obfuscation is \textit{currently} the more effective approach for medium and larger assignments, as fully generating only works well for smaller programs. Generated programs do not fulfill necessary functional requirements (not implementing the required behavior precisely) and even non-functional requirements like code style, thus requiring significant manual effort to improve them sufficiently.
Automatic obfuscation resembles human obfuscation practices, as a pre-existing solution is altered while \textit{trying} to preserve the program behavior.
For both approaches, the defense mechanisms have shown improved resilience.

For AI-generated solutions, there's an ongoing debate on whether this form of cheating qualifies as plagiarism~\cite{Novak2019, Saglam2024a}.
Our approach improves the detection rate by helping to recognize the similarities among generated solutions that occur due to the semi-deterministic nature of large language models.
This improvement is surprising, as the defense mechanisms are \textit{not} designed to detect AI-generated programs.

\subsubsection{On the Effectiveness of AI-based Obfuscation}

Our evaluation results show that the effectiveness of the defense mechanisms for AI-based obfuscation is less pronounced compared to their performance against algorithmic attacks. This can be attributed to two key factors.
First, the overall varying effectiveness of AI-based obfuscation plays a significant role. Our results indicate a strong variance in the similarity values achieved by AI-based obfuscation. While part of this variability can be explained by the different prompts used in our evaluation, this trend remains consistent even when examining the results for each prompt individually. For plagiarized programs that already exhibit a high degree of similarity to their original versions, there is limited potential and necessity for the defense mechanisms to increase that any further.
Second, generative AI employs a much broader range of modifications compared to algorithmic obfuscation techniques. Algorithmic methods typically rely on a well-defined, limited set of changes during obfuscation. Even refactoring-based obfuscation, which involves multiple refactoring operations, operates within a constrained set of transformations. In contrast, AI-based obfuscation introduces a far more diverse range of modifications, even when using the same prompt. In our evaluation, we observed strong variations in the types of changes applied by AI depending on both the prompt used and the dataset involved. These diverse modifications alter token sequences extensively, posing a challenge to the defense mechanisms.

Nonetheless, it is important to note that our evaluation still shows a notable improvement in resilience against AI-based obfuscation, even in the presence of these complex and varied changes. This demonstrates that while AI obfuscation is an effective technique, the defense mechanisms mitigate its effects.
Interestingly, the effectiveness of AI-based obfuscation attacks strongly varies depending on the dataset used. As illustrated in \autoref{fig:stage3-results}, AI-based obfuscation performs well for Homework-1, while it does not perform well for both PROGpedia datasets. TicTacToe and Homework-5 achieve mixed results. The median similarity differences range, depending on the dataset, between around ten and around 78 percentage points (see \autoref{tab:diff-gpt-obf}).
Although the evaluated plagiarism instances proved to be effective, the process of generating them was not straightforward.
In some cases, GPT-4 produces incomplete or invalid code. 
Despite over 50 attempts, we could not produce a valid result for three original programs, which all exceeded 300 LOC.
Thus, algorithmic obfuscation \textit{currently} exhibits more consistent results than AI-based obfuscation, and \textit{currently} can be just as effective.
However, AI-based obfuscation is more useful in avoiding detection during manual inspection, as it produces diverse modifications and can imitate human-made code.

\subsubsection{On the Effectiveness of AI-based Generation}

While AI-based generation works to a certain extent, its effectiveness is \textit{currently} limited. 
The programs generated entirely by GPT-4 did not fully comply with the specific requirements of the programming assignments, often resulting in additional output or slightly altered behavior.
These discrepancies suggest that fully AI-generated solutions may only be suitable for smaller, less complex assignments.
In our case, the TicTacToe dataset, with an average size of 236 lines of code, appears to be near the threshold where fully generated solutions start to exhibit these inconsistencies.

A noteworthy observation is that AI-generated programs are typically shorter than those created by human developers, especially within the TicTacToe dataset. This reduction in length may contribute to the higher degree of similarity observed between AI-generated solutions. While large language models like GPT-4 are not entirely deterministic, they exhibit a level of determinism sufficient for software plagiarism detection purposes. This inherent determinism, coupled with the more concise code produced by AI, may explain why AI-generated programs tend to resemble each other more closely than human-generated ones.
Finally, GPT-4 tends to produce placeholder comments instead of fully implementing certain methods, particularly when the task or method is not well-defined in the prompt. This behavior further limits the effectiveness of AI-based generation for complex assignments, as these incomplete implementations require additional manual intervention to complete.

\subsubsection{Emerging Threats}

While our results thus show that our contributions can effectively address \textit{current} threats of artificial intelligence, rapid advancements in this field may necessitate future re-evaluation. In the future, AI-based obfuscation methods may exhibit less variance in their effectiveness, thus increasing their reliability.
Similarly, new algorithmic attacks might emerge.
However, as discussed, all emerging attacks must affect the same attack surface, namely, the internal, linear program representation.
Thus, subsequence match merging will continue to provide resilience to emerging attacks. However, the degree of that resilience remains to be assessed.

The rapid development in the field of generative AI may lead to emerging threats that warrant close attention~\cite{Lancaster2023}. One area of particular concern is AI-generated programs. As generative AI advances, this might become feasible for larger programs and produce functionally correct programs for more complex assignments.
To detect such fully generated programs, detection systems need capabilities to detect obfuscation via implementation, which can be considered semantic clones. Here, caution is warranted. While matching full re-implementations seems desirable, it risks introducing significant false positives by flagging unrelated programs created independently by students. Note that unrelated solutions to a single problem can also be seen as semantic clones. Thus, we see the danger of creating unreliable detection systems, which may lead to unfairly penalizing students. Addressing re-implementation or semantic clones, therefore, raises philosophical questions about the boundaries of what type of plagiarism we actually want a detection system to target.

For fully generated programs, for example, via generative AI, plagiarism detection methods may not be sufficient for emerging attacks.
If traditional plagiarism detection methods, including the defense mechanisms evaluated in this paper, prove inadequate against more sophisticated AI-generated code, alternative techniques may need to be explored~\cite{karnalim2024}. One research area is the development of AI-based detectors that act as countermeasures to generative AI. However, at present, such AI-based detectors have not demonstrated sufficient reliability or performance, and they remain an area of ongoing research~\cite{WeberWulff2023, Pan2024, Khalil_Er_2023}. Another possibility lies in signature- or watermark-based methods, where the artifacts generated by AI are always identifiable as such. This approach would involve recognizing specific patterns or characteristics inherent to AI-generated content, allowing for consistent identification, regardless of the obfuscation techniques applied. Again, this is ongoing research~\cite{zhao2024provable, Jiang2023}. 
It is important to note, however, that these potential future developments lie beyond the scope of this paper and even outside of the research area of software plagiarism detection.

\subsection{Layering Defense Mechanisms}
Attack-specific defense mechanisms are highly effective, as they can be tailored with strong assumptions about specific obfuscation techniques in mind.
For their targeted obfuscation attacks, attack-specific mechanisms outperform attack-independent approaches.
This is evident in the case of token sequence normalization in \autoref{fig:stage1-results}, where the defense mechanism fully separates plagiarism pairs from original pairs, completely outperforming subsequence match merging.
However, attack-specific mechanisms mostly focus solely on a single known obfuscation attack type. Multiple attack-specific mechanisms must be combined to achieve broad resilience.
Additionally, attack-specific mechanisms can only be designed for known attacks and may not be equipped to handle emerging threats, as they rely on assumptions that may not hold true for unknown obfuscation techniques.
%%%
Attack-independent mechanisms, such as subsequence match merging, make fewer assumptions about the obfuscation techniques in use. Thus, they provide less resilience for a given obfuscation attack. 
Their strength, however, lies in providing broad resilience.
Throughout our evaluation, we observed that subsequence match merging consistently offered resilience across a variety of obfuscation attacks.
Because of its heuristic nature and the fact that it operates at a high level of abstraction, it can provide \textit{some} resilience against unknown and emerging obfuscation attacks.
Attack-independent approaches are essential for defending against emerging threats.
Since they make fewer assumptions about the nature of incoming obfuscation attacks, they offer a level of protection against unknown attacks that attack-specific mechanisms may not.

The ideal solution is to combine multiple defense mechanisms, leveraging both attack-specific and attack-independent defense mechanisms.
This strategy provides targeted resilience against well-known or highly effective obfuscation attacks while also offering broad protection against unknown or emerging techniques.
Layering multiple defenses is a well-established strategy in information security and risk assessment, often referred to as the \textit{Swiss cheese model}~\cite{Reason1990} or \textit{defense in depth}~\cite{Stytz2004, Lippmann2006, Anderson2020}.
When using this layered approach, it is critical to ensure compatibility between defense mechanisms to avoid unintended side effects that could reduce overall obfuscation resilience or detection quality.
In the context of software plagiarism detection, it is beneficial to allow users to enable or disable different defense mechanisms depending on their needs or to mitigate potential side effects.
The defense mechanisms are designed to be minimally intrusive, enabling them to be layered with other approaches.
In our evaluation, we examine the combination of defense mechanisms to check for adverse side effects.
While in some cases, individual mechanisms outperformed the combination, the difference from the second-best mechanism is always negligible.
Therefore, our mechanisms can be safely used in a layered defense strategy.

\section{Related Work}\label{sec:eight}
This section discusses research from areas intersecting with this paper’s contributions.

% --------------------------------------------------------------------------------------------
\subsection{Software Plagiarism Detection Systems}
Despite its early roots~\cite{Ottenstein1976}, research in software plagiarism detection has seen a resurgence in recent years~\citet{Novak2019}.
Most software plagiarism detection approaches compare the structure of the code~\cite{Nichols2019, Novak2019}; among them, token-based approaches are the most popular tools employed in practice.
JPlag~\cite{prechelt2000} and MOSS~\cite{MOSS} are the most widely used tools \cite{Aniceto2021, Novak2019}.
Furthermore, JPlag is most frequently referenced and compared to \cite{Novak2019}.
Other tools mentioned frequently are Sherlock~\cite{Joy1999} and SIM~\cite{SIM}. However, they are partially outdated or no longer maintained.
Dolos~\cite{Maertens2022} is a more recent tool inspired by MOSS and JPlag but currently supports only single-file programs, which limits its applicability.
All mentioned approaches are token-based and find matching fragments via hashing and tiling~\cite{prechelt2002, MOSS}.
Some recent approaches also employ machine learning for plagiarism detection~\cite{Ebrahim2024}.
We use JPlag as a baseline in our research; however, the approaches extend to any token-based detector and can be generalized to structure-based methods.
\textit{The mentioned works evaluate their tools with manually-obfuscated plagiarism. 
We specifically focus on evaluating automated obfuscation.}

% --------------------------------------------------------------------------------------------
\subsection{Obfuscation Attacks and Their Mitigation}
Obfuscation attacks present a significant challenge for software plagiarism detection. While obfuscation has long been a concern~\cite {zhang2014b, Karnalim2016, Novak2019}, research on defending existing state-of-the-art plagiarism detection tools from automated obfuscation is limited. Most recent studies focus on developing entirely new detection systems that often remain inaccessible to the public, as noted by \citet{Novak2019}. Research on mitigating obfuscation usually focuses on manual obfuscation. In the following, we discuss notable exceptions.
\citet{DevoreMcDonald2020} introduce \mossad, a tool that uses genetic programming techniques to automatically generate semantically equivalent but undetectable plagiarized code variants, defeating detectors like Moss and JPlag. Its non-deterministic transformations mimic authentic student submissions.
\citet{Biderman2022} show that language models like \textit{GPT-J} can produce correct, syntactically diverse solutions that evade Moss detection with minimal human input, raising concerns for academic integrity as AI tools become more accessible.
Similarly, \citet{Karnalim2020} evaluate 16 preprocessing techniques for source code similarity detection, finding that methods like identifier removal and syntax tree linearization improve detection effectiveness. However, such techniques offer limited resilience against broader obfuscation strategies.
\textit{These works typically focus on individual obfuscation attacks. In contrast, our work addresses a broader spectrum of automated obfuscation strategies and shifts the focus from attack feasibility to evaluating concrete defense mechanisms.}

\subsection{Generative AI in Programming Education}

\citet{chen2024b} investigate the impact of generative AI on academic integrity in an introductory programming course.
They show that suspected plagiarism increased and shifted from traditional sources to AI tools.
The results of their regression suggests that increased plagiarism may lead to decreased learning outcomes.
In contrast to these results, other studies observe no difference.
\citet{Xue2024} conduct a controlled study on ChatGPT’s impact in CS1 programming education with 56 participants.
Their results showed no significant difference in learning outcomes between groups. 
Most students held neutral views but expressed concerns about ethical issues and ChatGPT’s inconsistent results. 
\citet{Choudhuri2024} explore the impact of conversational generative AI on supporting students in software engineering tasks. Their study with 22 participants found no significant difference in productivity or self-efficacy compared to traditional resources, but noted significantly higher frustration levels. 
%%%%
\citet{Karnalim2025} investigate student perceptions of AI-assisted plagiarism in programming education by comparing it to traditional plagiarism scenarios. Based on survey responses from 66 introductory and intermediate programming students, the study finds that students view AI assistance as morally comparable to help from peers.
The study suggests that student awareness and interpretation of AI-assisted plagiarism vary by experience level.
\citet{Cipriano2024} examine the performance of large language models in object-oriented programming (OOP) exercises. Using real-world educational tasks and automatic assessment tools, they found that while LLMs often produce working solutions, they frequently neglect OOP best practices. The study highlights the need to emphasize code quality in programming education.
\textit{Generative AI is becoming an integral part of programming education, and as educators, we have to deal with its impact. For this reason, this paper specifically investigates AI-based obfuscation.}

\subsection{Detecting AI-Generated code}
\citet{karnalim2024} propose a lightweight AI-assisted code detector based on code anomaly features, which uses 34 features spanning various program elements to identify unusual patterns that may indicate AI assistance. Evaluated across three datasets, their approach shows promising results. However, the detection effectiveness drops significantly when students collaborate or use AI only partially.
\citet{Orenstrakh2024} evaluate the effectiveness of eight publicly available detectors for identifying LLM-generated content. They collected 124 human-written student submissions from before ChatGPT and compared them with 40 ChatGPT-generated samples. They find that detection accuracy significantly drops for programming code, non-English text, and content modified with paraphrasing tools, highlighting current limitations of such detectors.
Similarly, \citet{Suh2025} investigate the challenge of detecting AI-generated code, noting that current detection tools perform poorly and lack generalizability. To address this, they propose enhanced approaches such as fine-tuning LLMs and using machine learning classifiers with static code metrics or AST-based embeddings. Their best model outperforms GPTSniffer, achieving an F1 score of 82.55.
Moreover, \citet{Pan2024} examine the ability of ChatGPT to evade detectors for AI-generated content in programming education. Using a dataset of 5,069 human-written Python solutions, they prompted ChatGPT with 13 code problem variants and evaluated five detectors. Results show that current detectors struggle to distinguish AI-generated code from human-written code reliably.
\textit{Given that AI code detectors are unreliable, it is especially relevant that software plagiarism detectors can produce suspiciously high similarity values for AI-generated programs produced by the same model. We analyze this effect in our evaluation.}

% --------------------------------------------------------------------------------------------
\subsection{Clone Detection}\label{sec:rw-clone-detection}
Reusing source code via copying commonly leads to code clones~\cite{Roy2009}, which impedes modern software development~\cite{juergens2009}.
Code clones are created accidentally~\cite{juergens2009}, while plagiarism is a deliberate act.
While both clone detection and plagiarism detection are software similarity problems~\cite{Novak2019}, they ultimately differ in many aspects~\cite{mariani2012}.
In contrast, code clone detection does not consider scenarios where an adversary attempts to affect the process, as code clones typically arise inadvertently~\cite{juergens2009}. As a consequence, clone detectors are vulnerable to obfuscation attacks.
Plagiarism detection approaches must deal with an additional layer of complexity introduced by the adversary-defender-scenario~\cite{Saglam2024b}.
Still, many clone detection approaches share similarities in their employed techniques~\cite{wang2018}. 
\textit{In summary, while clone detection is a related field, these works are not applicable for automated obfuscation.}

\section{Conclusion}\label{sec:nine}

In this paper, we analyze the state of plagiarism detection by evaluating defense mechanisms for plagiarism detection systems that provide resilience against automated obfuscation attacks.
The defense mechanisms include both a tailored approach that targets specific obfuscation types and a broad, attack-independent mechanism that ensures resilience against a wide range of obfuscation strategies, including emerging and unknown threats.
Our evaluation shows that the defense mechanisms demonstrate significant improvements in detection rates across a wide variety of obfuscation methods, achieving significant gains in similarity scores even under AI-based obfuscation.
In conclusion, this paper offers key insights into emerging challenges in academic integrity. By contributing empirical insights and discussion about the current state of plagiarism detection, we support educators in maintaining integrity across programming assignments in a landscape increasingly shaped by automated obfuscation~\cite{Foltynek2020, Biderman2022}.

Despite these advancements, it would be inadvisable to consider the problem of automated obfuscation in software plagiarism as solved.
The rapid development of generative AI continues to reshape education. 
Educators and researchers must learn to adapt to these developments to anticipate possible implications on academic integrity.
Addressing AI-based obfuscation attacks will require even greater focus in future investigations.
Nevertheless, artificial intelligence should not be viewed as an all-encompassing threat; instead, it creates new challenges \textit{and} opportunities. Critically, we should focus on understanding its limitations and leveraging its strengths~\cite{Saglam2024Keynote}. 
At the same time, it is essential to complement technical efforts with an open educational and ethical discourse on what constitutes plagiarism in the age of AI. Detection tools can identify suspicious patterns, but they cannot define the boundaries of misconduct -- this remains a fundamentally human and pedagogical responsibility.

Plagiarism detection systems excel at identifying structural similarities, while humans are uniquely capable of quickly identifying semantic nuances between programs given these similarities. This combination of large-scale analysis and human inspection has thus been very effective and will remain so for the foreseeable future. Fundamentally, however, tool-based plagiarism detection must always involve a human decision as a final step.
In addition to detection, prevention is just as important in combating plagiarism~\cite{Simon2016}. Students often resort to plagiarism when overwhelmed and believe they have no other options~\cite{Amigud2019}.
Proactive strategies such as student training, thoughtful assessment design, clear institutional policies, and counseling should thus be combined with detection approaches~\cite{Lancaster2023}.

\begin{acknowledgements}
This work is supported by the CHAINS project funded by the Swedish Foundation for Strategic Research (SSF). 
\end{acknowledgements}

\section*{Declarations}
The authors declare that they have no conflict of interest. All data and code related to this paper is available at \cite{replication-package}.

\bibliographystyle{plainnat}
\small
\bibliography{bibliography}

\end{document}